\documentclass[prd, aps, superscriptaddress, showpacs]{revtex4-1}

\usepackage{amssymb,amsmath,amsthm,graphicx,amscd}
\usepackage[mathscr]{eucal}
\usepackage{enumerate,color,verbatim,multirow,comment,array,ulem,chngcntr,caption}
\usepackage{bm,pbox}

\newcolumntype{C}[1]{>{\centering\arraybackslash}m{#1}}
\newcolumntype{L}[1]{>{\raggedright\arraybackslash}m{#1}}
\newcolumntype{N}{@{}m{0pt}@{}}
\counterwithin{paragraph}{section}

\begin{document}

\title{Distance and Coupling Dependence of Entanglement \\ in the Presence of a Quantum Field  }  

\author{J.-T. Hsiang}
\email{cosmology@gmail.com}
\affiliation{Department of Physics and Center for Field Theory and Particle Physics,\\
Fudan University, Shanghai 200433, China }
\author{B. L. Hu}
\email{blhu@umd.edu} 
\affiliation{Department of Physics and Center for Field Theory and Particle Physics,\\
Fudan University, Shanghai 200433, China }
\affiliation{Joint Quantum Institute and Maryland Center for Fundamental Physics , 
University of Maryland, College Park, Maryland 20742-4111, USA}

\begin{abstract}
We study the entanglement between two coupled detectors,  whose internal degrees of freedom are modeled by harmonic oscillators,  interacting with a common quantum field, paying special attention to two less studied yet important features: finite separation and direct coupling. Distance dependence is essential  in quantum teleportation and relativistic quantum information considerations. The presence of a quantum field  as the environment accords an indirect interaction between the two oscillators at finite separation of a non-Markovian nature which competes with the direct coupling between them.  The interplay between these two factors results in a rich variety of interesting entanglement behaviors at late times.  We show that the entanglement behavior reported in prior work assuming no separation between the detectors can at best be a transient effect at very short times,  and claims that such behaviors represent  late time entanglement are misplaced.     Entanglement between the detectors with direct coupling  enters in the consideration of  macroscopic quantum phenomena and other frontline issues.  We find that with direct coupling entanglement between the two detectors can sustain over a finite distance, in contrast to the no-direct coupling case reported before,  where entanglement can not survive at separation more than a few inverse high frequency cutoff scales.  This work provides a  solid platform necessary for further systematic investigations into the entanglement behavior of continuous variable quantum systems. 
\end{abstract}

\pacs{
03.65.Ud, 
03.65.Yz, 
03.67.-a}, 

\maketitle

\baselineskip=18pt
\allowdisplaybreaks

\section{Introduction}
As the uniquely distinguished feature of quantum  mechanics \cite{Schrodinger}, and,  as the  primary resource of quantum information processing (QIP) \cite{Nielsen} a deeper understanding of entanglement carry both foundational  \cite{Peres} and practical values, such as the application to quantum teleportation \cite{QTelepLight}.  The first major task of finding mathematically sound \cite{QInfoMathBooks} definitions to quantify the existence of bipartite and multipartite entanglement  for discrete variable quantum systems such as `concurrence' \cite{Wooters} and continous variable quantum systems such as `negativity'  \cite{EntCriteriaCV} was undertaken in the 90s \cite{QInfoCVBooks,QInfoCVReview}.  From physical considerations    a system where quantum information processing is carried out is always under the influence of its environments, whose effects can be represented as noises of different nature which could be detrimental to QIP.   Environment-induced quantum decoherence and disentanglement are two  vital  processes which need be accounted for and understood  well to enable one to find ways to mitigate or manage them.  This second major task is most effectively undertaken with the methods of quantum open systems \cite{qos}  based on various aspects of the quantum Brownian models.   Environments  when  suitably prepared or  attuned to can also assist in maintaining or even generating entanglement \cite{Braun,Knight}.   Serious studies of entanglement between two qubits \cite{EntQB,ASH,Kanu,nMark2LA} and continous variables  \cite{EntCV,Goan,HB08,Paz,LH09,Zell09} appeared in the last decade,   Our studies on how entanglement in a quantum system is affected by its environment follow this call,  relying mainly on functional methods \cite{FeyVer,CalLeg,HPZ,HM94}, and in the case of  Gaussian systems \cite{Gaussian,Charis,ChrisNQBM},  the covariance matrix  theory .   

In this paper we consider the entanglement between  two harmonic oscillators,  call detectors \footnote{A detector refers to a physical entity with internal degrees of freedom (idf) , such as a real (not the idealized two level) atom.}, which make up our system, interacting with a quantum scalar field acting as its environment \cite{UnrZur89,CalHu89} with special attention paid to two factors:  distance and coupling dependence of the entanglement dynamics. 
 
One important factor which has been largely ignored or grossed over is how entanglement between two qubits (representing discrete variables) or harmonic oscillators (representing continuous variables) depends on their spatial separation.  This is one of the two primary aims of our prior studies \cite{ASH,LH09}.   Distance dependence is recognized as an essential factor in relativistic quantum information \cite{RQI} such as  entering in quantum teleportaion considerations \cite{LCHQTelep}. Even with stationary detectors this is a relevant factor,  because the two detectors are linked by a field, and information propagates through the field as their medium, whereby entanglement can be induced, even generated.  Many prior work  \cite{Goan,HB08,Paz} assumes  zero spatial separation between  the two detectors and report on  the (common) environment-induced   entanglement dynamics --  such as generation, death and revival.   One may casually think that  the distance effect is  not so important because at most it invokes a rather weak retardation effect from the field environment, or a small contribution from the bath correlations.  But a moment's thought calls into question the following:   If there is no direct coupling, as assumed, how could two detectors get entangled via a field which acts  only at the same point where the two detectors are located?  Both detectors would read off the same local value of the field, with no dynamics of the field involved.  The  environment would affect each detector's entanglement dynamics but would not enter into influencing the combined two-detector system over and beyond the simple sum of each.  Equivalently this amounts to a contact potential with instantaneous action of the field. Since this is the only means of interaction between the two detectors,  the zero separation is clipping the wings of the field and thus the setup has some intrinsic deficiencies.  

For an oscillator bath, one may argue that even without spatial correlation there should still be time-time  autocorrelation as influenced by the environment (via the self-induced non-Markovian dissipation kernel).  But the fact is, each detector would get the same influence from the environment, as it interacts with the environment in the same way as the other does. Without spatial separation or without direct coupling this would be like two separate detectors each conducting its own business albeit interacting with the same environment.  The environment certainly influences each detector,  but it does not contribute to any additional inter-oscillator entanglement beyond which it imparts to each detector.  For these reasons,  conclusions drawn from calculations making this assumption can only be valid for a very short time scale, reflecting a transient behavior which cannot be extended to late times.  Thus the meaning conveyed by the word `common' in  ``entanglement induced by a common bath" implying that there is added inter-detector entanglement due to the fact that the two detectors share a common bath could be misleading.   The proper treatment even for the zero separation case would have to begin with a finite separation, give full dynamics to the field,   allow it to partake of the dynamics of the combined two-detector system,  then take the small separation limit.   Indeed in our present calculation we can pinpoint and expose this shadowy feature and show the cause of it explicitly.  This is one exemplary observation which calls for a careful reconsideration of prior results based on the zero-separation condition.

The other important factor is allowing for direct coupling between the  two detectors.  This  enlarges the scope from cases with no direct coupling,  such as in neutral atoms,  to include a wider range of inter-detector interactions.   The type of coupling we treat here in detail is of the quadratic type between the detector's internal degrees of freedom (denoted as $\chi$ here, or $Q$ in our earlier papers \cite{RHA,LCH08,LH09}), which is of the same form where the HPZ-type of master equation \cite{HPZ,HalYu} was earlier derived for two coupled oscillators  \cite{CYH08}.   By keeping the whole system Gaussian, we are able to produce exact analytical solutions valid even for strong couplings.  Our present work is a direct descendent from and a generalization of our earlier work \cite{LH09} on the temporal and spatial dependence of entanglement with the same setup,  but now with non-zero inter-detector coupling.  This allows us to see both a) the competition between the direct coupling in protecting entanglement and the environmental influence of corrupting entanglement and b) the direct coupling effect versus the indirect or induced effect of the field on the entanglement  between the detectors,  the latter came alive from a  finite separation.   

We add that these considerations are not just adding details to established prior work, they surely do, but more significantly they enable us to identify some  largely unnoticed blind spots in earlier treatments, to rectify  possible misleading conclusions affected by them,  and to provide a more reliable foundation for further developments.  Including these two essential factors -- distance and coupling -- in the consideration of the entanglement dynamics of this workhorse model for continuous variable quantum systems and its extension could provide a more complete, more correct and  more accurate analysis for many problems it can tackel with,  such as expounding some quantum foundation issues (e.g., \cite{KLRA15}),  quantum  information and teleportation applications (e.g., \cite{ChiAdePRL14,LCHQTelep}),  understanding macroscopic quantum phenomena (e.g., \cite{CHS13,HZH13}) and exploring the interesting new field of quantum thermodynamics (e.g., \cite{Galve,ValAloKoh,ValCorAlo,HHQEnt1,HHQEnt2}).

Below we first  provide the background by describing some representative prior work in their respective ranges and the claimed results. We then provide a qualitative description of the main features of this system,  present a brief summary of our findings and note the main differences from prior work.\\

\subsection{Prior Works} 

To provide the backdrop for this investigation we tabulate what has been done before in  five oft-cited papers: Columns in Table~\ref{T:ekfdkww} indicate the authors  corresponding to References \cite{Goan,HB08,Paz,LH09,Zell09} respectively;  Rows A-H indicate whether distance and coupling between the detectors are considered,  properties of the bath, etc.   The footnotes beneath the Table signify some special conditions related to a specific feature.  A quick glance of Items A and B indicate that the two papers closest to our concern are that of \cite{Paz} and \cite{LH09}, factoring in much overlap of contents in \cite{Zell09} with \cite{LH09}. 

\begin{table}
\begin{center}
  \begin{tabular}{ | L{3cm} ||C{2.7cm}| C{2.7cm} | C{2.7cm} | C{2.7cm} | C{2.7cm} | N}
    \hline
     Comparison Table${}^{(1)}$	& Liu-Goan \cite{Goan} & H\"orhammer-B\"uttner \cite{HB08} & Paz-Roncaglia \cite{Paz}& Lin-Hu \cite{LH09}& Zell-Queisser-Klesse \cite{Zell09} &\\ [20pt]\hline
    A. Spatial separation 	& N & N & N & Y & Y &\\ [20pt] \hline
    B. Direct coupling between detectors    	& N & N & Y${}^{(2)}$ & N & N &\\ [20pt] \hline
    C. Bath type			& $N$-oscillator & $N$-oscillator${}^{(3)}$ & $N$-oscillator & scalar field & Extended $N$-oscillator${}^{(4)}$ &\\ [20pt] \hline 
    D Time regime			& Transient & Transient & Transient \& late time & Transient \& late time & Transient \& late time &\\ [20pt] \hline
    E. Equation of motion	& Master Eq. with RWA+Markov approximation & Exact master equation & Exact master equation & Quantum Langevin Eq. & Quantum Langevin Eq. &\\ [20pt] \hline
    F. Initial state		& Two-mode squeezed vacuum & Two-mode squeezed state & Two-mode squeezed state & Two-mode squeezed state & ground state &\\ [20pt] \hline
    G. Computation approach& Numerical & Numerical & Analytical \& numerical & Analytical & Numerical &\\ [20pt] \hline 
    H. Critical parameter & None & None & $\beta_{c}=\dfrac{1}{\sqrt{2m\omega}}$ ${}^{(5)}$ & $d_{c}\sim\dfrac{\pi}{\omega\Lambda_{1}}$ ${}^{(6)}$ & $d_{c}\sim\Lambda^{-1}$ &\\ [20pt] \hline
  \end{tabular}
\end{center}
	\captionsetup{singlelinecheck=off,justification=justified}
\caption[blablabla]{%
Comparison of five oft-cited papers.\\
\begin{enumerate}
	\item Symbols:  $m$, $\omega$ and $\gamma$ denote the mass, natural frequency, and damping constant of both oscillators, respectively. The parameter $\ell$ describes the spatial separation, and $\sigma$ represents the coupling strength between the two oscillators.  
	\item A direct coupling constant is present in the formalism but was renormalized to zero in their  calculation.    
	\item The authors assume that the dynamics of the relative variable comes about as the solution of a  Born-Markovian master equation for a damped oscillator while in actuality the relative variable really behaves like a free oscillator, a fact from symmetry considerations.
	\item Spatial dependence is inserted in the interaction term of the system oscillators and the bath oscillators, so in a sense the configuration  is very similar to that in~\cite{LH09}
	\item The critical temperature is found for the coherent state with direct coupling strength renormalized to zero so it is independent of the squeezing parameters and the direct coupling. We will treat thermal entanglement in a sequel paper \cite{HHQEnt1}.
	\item The dimensionless parameter $\Lambda_{1}$ is related to the resolution $\Delta t$ of the detector by $\Lambda_{1}=-\ln\omega\,\Delta t-\gamma_{\epsilon}$. The critical separation is independent of the coupling constant only because it is obtained in the vanishingly weak coupling limit.
\end{enumerate}
}\label{T:ekfdkww}
\end{table}

\subsection{Direct and induced interactions, Symmetry Considerations, Separation and Coupling Dependences }

We assume that two detectors stay at rest at a finite distance $\ell$ apart in a common bath modeled by a massless scalar field. The internal degree of freedom $\chi$ of each detector is described by a harmonic oscillator, which is directly coupled to the internal degrees of freedom of the other oscillator and interacts with the environmental scalar field, both bilinearly.  To  assess the validity of the results reported in prior and the present work it is necessary to understand how different interactions in the system affect the dynamics and how its behavior varies in the different parameter regimes of interest, as we now describe.  

\paragraph{Induced interaction and non-Markovian Dynamics}
 
The two types of interactions between the internal degrees of freedom of the two detectors,  a direct inter-oscillator coupling, and  an indirect, nonlocal, retardation interaction, mediated by the common environment field determines the dynamics of the system quite differently.  We can see why the indirect interaction produces non-Markovian dynamics as such:  the interaction of one oscillator with the field produces some disturbance in the field (or signal) which  takes a finite amount of time (follows the lightcone) to propagate to the other oscillator located at some distance {and modify its motion}.  The second oscillator does the same in emitting its signal while receiving the signal from the first oscillator, but at a later time.   The history-dependent nature is what gives the induced interaction a non-Markovian character. This is explained and well-illustrated in the results of our prior work \cite{LH09}.

We may add a few words on the difference in the role of the environment made up of a scalar field (e.g., \cite{UnrZur89,CalHu89} ) versus  $N$-harmonic oscillators (e.g., \cite{FeyVer,CalLeg}.  To see this relation explicitly, as  e.g., in \cite{HM94}, where a scalar field is explicitly expressed in terms of parametric oscillator modes.  But there are subtle differences:  We mention just two here.  First, the spectral density of a field is fixed, and takes on a different form for different dimensions.  However, one can specify the form of the spectral density function for an $N$-oscillator bath.  Second, how to deal with a divergent contribution from the high frequency end of the spectrum.  There are well-justified ways to obtain a well-defined integral over the whole frequency domain in the case of a quantum field bath, by way of regularization or renormalization methods. For an $N$-oscillator bath the spectral density is often crafted to take on a form that its value tapers off to zero when the frequency is greater than a certain high frequency cutoff scale. As a consequence, in the time domain the dissipation kernel will subside at a rate of the inverse cutoff scale. For a very high frequency cutoff this falloff is very steep, but for not so high a frequency cutoff,  reflective of some `softer' physical environment,  the dissipation  kernel would contain a feature which resembles a self-induced time-nonlocal interaction of each oscillator with itself.  This feature is usually ignored  under the assumption that fewer high-frequency modes in the environment take part in the interaction with the subsystems,  from unspecified physical considerations.  

\paragraph {Symmetry and Dynamical Considerations}

One can regroup the two internal degrees of freedom of the two detectors into a symmetric (or center of mass) variable and an anti-symmetric (relative or difference) variable and use them to describe the  dynamics of the combined (two-oscillator) system.  The dynamics of these two variables (interchangeably may be called `modes')  are very different and dictate very different entanglement behavior. (An explicit illustration of how symmetry determines the entanglement behavior is given in \cite{Kanu} for a system of two qubits interacting with two cavity modes.)  In a $N$-oscillator bath the relative mode is found to be freely evolving, totally decoupled from the bath.  The decoupled relative mode dynamics is the root cause for the fact that the late-time entanglement of the combined system oscillates in time and shows dependence on the initial condition of the joint system,  even  though it is acted upon by a bath whose damping function one expects to wash out the sensitive dependence of the initial conditions.   This entanglement behavior is unavoidable in the dynamics of two oscillators at zero-separation (attempts to remove this behavior by adding some damping by hand is unwarranted and unphysical).   However, with finite separation between the two detectors, field-induced effects set in to provide a natural damping effect on the relative or anti-symmetric mode dynamics.  From a Taylor expansion of the retardation terms we see immediately that the entanglement dynamics of the two oscillators at the same location shows up at very short times as transients. Thus our earlier comment that results obtained from the zero-separation setup cannot be taken to represent the long time behavior.   Including higher-order terms in this expansion shows  that the relative mode decays with time at a slow rate.  For  sufficiently long time after the joint system has reached full relaxation,  the entanglement approaches a constant and becomes  independent of the initial conditions.

\paragraph{Oscillator-Bath Coupling and Instability}

{For non-Markvian dynamics in the current configuration, strong oscillator-environment coupling will not result in an overdamped motion as it usually does for Markovian dynamics. Instead it tends to induce instability. This in a sense is similar to parametric oscillation: the disturbance from  one oscillator can have a secular amplification effect on the motion of the other oscillator.}

More details can be found in the calculations described in the following sections.

\subsection{Temporal, spatial and coupling strength- marked regimes}

{In the following, we assume the two oscillators have the same natural frequency $\omega$ and damping constant $\gamma$.  The parameter $\ell$ describes the spatial separation, and $\sigma$ represents the coupling strength between the two oscillators.}

\begin{itemize}
		\item \textbf{Short/long Time}.   The dividing time scale is  the relaxation time $\gamma^{-1}$.   We are mostly interested in the entanglement behavior of the two coupled oscillator system at late times, after it has fully relaxed or reached equilibrium with the environment. We assume that the oscillators have exchanged their (non-Markovian) mutual influence via the field many times, which implies that $\gamma^{-1}>\ell$. For a small  $\gamma/\omega^{2}\ell$ factor, the non-Markovian effects decays rapidly after a few exchanges.
		
	\item \textbf{Short/large Separation}.  When the distance between the two oscillators is shorter than $\gamma/\omega^{2}$, either the perturbation calculation is unreliable or the motion of the joint system becomes unstable. The instability is a feature of non-Markovianity different from the simplifying (but often lack of justification) assumption of a Markovian process leading to overdamped motion of the system.  Another interesting scale is $\gamma/\sigma$ where  a  change in the entanglement measure occurs. It is the value where the direct interaction (inter-oscillator coupling) is comparable in strength with the indirect (field-induced) non-Markovian effect. When the separation between the oscillators is larger than this scale, the direct coupling between the oscillators dominates the dynamics. On the other hand, when the separation is shorter than that scale, the non-Markovian (field-induced) effect is the determining factor in the motion of the joint system. Other than these two length scales, we refer to $\omega\ell\gg1$ as large separation and $\omega\ell\ll1$ as small separation.
	
	\item \textbf{Weak/strong Coupling}.   For the coupling between the bath and the oscillators, we only consider the case $\gamma/\omega<1$, that is, weak coupling. When $\gamma$ is comparable or greater than $\omega$,  instability of the motion tends  to occur unless $\omega\ell\gg\gamma/\omega$. This range is less interesting because it corresponds to the very large separation cases. For direct couplings between the oscillators, we only require that the resonance frequencies of the normal modes be real. This implies $\sigma<\omega^{2}-\gamma^{2}$, which is not a stringent constraint. Thus $\sigma$ can be comparable with $\omega^{2}$ and is considered strong.   Thus we may refer to $\sigma<2\gamma/\ell$ as weak coupling because the direct coupling plays a minor role compared with the non-Markovian indirect interaction. In certain cases, we may encounter $\sigma<\gamma\omega<2\gamma/\ell$. It corresponds to vanishing inter-oscillator coupling. This is equivalent to the cases, in the table, where the inter-oscillator coupling is not present.
	
\end{itemize}	

\subsection{Late time entanglement behavior}

{For  finite separation between the two detectors,  due to the field-mediated  (non-Markovian) mutual influence interaction, the effective damping constant and the effective oscillating frequency are separation-dependent and do not take on the same values between the two (symmetric vs antisymmetric) normal modes.  Being at late times, the motion of either normal mode is fully relaxed or has reached equilibrium. Since the strength of mutual influence  depends on the separation between the two oscillators,  and there is  competition between the direct inter-oscillator coupling and the indirect, environment-mediated interaction,  the late time entanglement between two oscillators can have quite distinct behaviors, depending on the ratio $\varsigma \equiv \dfrac{\sigma\ell}{2\gamma}\gtrless1$, as follows:} 

\begin{itemize}
	\item When $\varsigma>1$: this is the regime where the direct coupling wins over the indirect interaction between the oscillators. We find that the late time entanglement can improve with 
		\begin{itemize}
			\item stronger inter-oscillator coupling,
			\item larger oscillator separation, and
			\item weaker oscillator-bath interaction.
		\end{itemize}
	\item When $\varsigma<1$: the effects of direct coupling is subdominant, compared with the field-induced interaction. We find that the late-time entanglement is favored with
		\begin{itemize}
			\item weaker inter-oscillator coupling,
			\item shorter oscillator separation, and
			\item stronger oscillator-bath interaction.
		\end{itemize}
\end{itemize} 
Alternatively, heuristically speaking, for sufficiently strong inter-oscillator coupling, their late time entanglement can survive even when they are very far apart. The late time entanglement worsens if the two oscillators are originally placed closer. When their separation is less than a critical value {$\ell_{>}$, as shown in \eqref{E:oqanpret} or \eqref{E:lkiqnkzaq}}, the oscillators will instead become separable at late time. It means that direct coupling is not enough to overcome the effect of both the field-induced non-Markovian interaction effect and the local vacuum fluctuations of the field. When we place the oscillators at a separation even shorter than another critical value {$\ell_{<}$  as in \eqref{E:dislwsd}}, the entanglement between the oscillators can again be sustained at late time. Now that the non-Markovian processes have dominated over those from direct coupling the results can be less intuitive and harder to interpret. If we set up oscillators at shorter and shorter separation, we find that their motion eventually become unstable.  Quantitative results are contained in the last section.

In short we see that direct coupling can play a key role in maintaining entanglement between the subsystems over finite distance. The type of inter-oscillator coupling we used may be idealized in the sense that its strength remains constant over large separation.  Even when the inter-oscillator coupling is separation-dependent and say falls off relatively slowly with larger separation, the arguments above should still  be valid over the region where the inter-oscillator coupling is much stronger than the mutual-influence interaction between the oscillators.

The organization of this paper is as follows:  In Sec. 2 we discuss the dynamics of two coupled quantum oscillators at finite separation as the system in a common quantum field environment, paying particular attention to their late-time evolutions.  In Sec. 3, we construct the covariance matrix from the uncertainties or the cross-correlation of the canonical variables of these two oscillators.  This enables us to compute the entanglement measure. Due to a multitude of scales, in Sec. 4 we highlight some constraints on the choice of the parameters, which will facilitate the discussions of entanglement measure. In Sec. 5 we introduce negativity as a quantifiable entanglement measure for the symmetric Gaussian state, and overview its qualitative behaviors in the current configuration. In Sec. 6, we give a quantitative analysis of late time entanglement of the coupled oscillators in the common bath. Finally we summarize our findings and their physical implications in Sec. 7. Details of key calculations of the covariance matrix elements are shown in the Appendix~\ref{E:nvndfkse}.


\section{Two detectors at  finite separation in a common field environment}

We consider a generic setup for the investigation of quantum entanglement between continuous variables directly coupled with each other while both interacting  with the same environment.  Consider two detectors $S_{1,2}$  located at $x=\pm\ell/2$ which remain at rest throughout their dynamics.  The internal degrees of freedom (idf) of  $S_{1,2}$ are represented by harmonic oscillators $\chi_{1,2}$ respectively.  In a realistic setting the detectors may represent (harmonic) atoms  whose electronic energy levels are their idfs.   We  assume the idfs $\chi_{1,2}$ interact directly with each other with a bilinear coupling. In addition,  each idf interacts with a common environmental quantum field $\phi$ at the locations of the detectors. The quantum field provides an indirect interaction  between $\chi_{1,2}$  which is non-Markovian in nature,  in the following sense:  Any activity in the idf of one detector will generate a perturbance of  the field which is picked up by the  idf of the other detector.  Because the propagation of signals in a field cannot be instantaneous, the finite time required engenders retardation effects  and the evolution histories of the detectors are intertwined imbued with memory. This is what gives rise to the non-Markovianity in the  evolution of the idf of the system of two detectors, as can be seen more clearly in the same setup without direct coupling as from the results of \cite{LH09}. 

Therefore  in this configuration three different types of correlations need be considered  in the two detectors-quantum field system. The first type of correlations comes from the direct interaction between the oscillators'  internal degrees of freedom, and is not directly related to the environment. The other two types are directly related to the environment: one results from the preexisting correlations of the environment field at locations of the detectors and the other is due to the retarded propagation of the field as a consequence of its interaction with the oscillators.

\subsection{Configuration}
The action that describes the present configuration is given by
\begin{align}
	S[\chi,\phi]&=\int\!ds\;\biggl[\sum_{i=1}^{2}\frac{m}{2}\,\dot{\chi}_{i}^{2}(s)-\frac{m\omega_{b}^{2}}{2}\,\chi_{i}^{2}(s)\biggr]-\int\!ds\;m\sigma\,\chi_{1}(s)\chi_{2}(s)+\int\!d^{4}x\;j(x)\phi(x)+\int\!d^{4}x\;\frac{1}{2}\,\partial_{\mu}\phi\partial^{\mu}\phi\,,
\end{align}
with $x=(t,\mathbf{x})$ the spacetime coordinate. The internal degree of freedom $\chi_{i}$ of the $i^{\text{th}}$ detector ($i=1,2$) is modeled as a harmonic oscillator with equal mass $m$ and equal (bare) oscillating frequency $\omega_{b}$;   $\sigma$ is  the coupling strength for the direct coupling between $\chi_{i}$ and $e$  is the interaction strength between $\chi_{i}$ and $\phi$. Thus the only feature that distinguishes between the two detectors is their spatial locations $\mathbf{z}_{i}$ . Assuming a finite separation between them enables us to see the full dynamics of environment-induced entanglement which is absent in most prior work  on continuous variable dynamics (with the exception of \cite{LH09}). The current $j(x)$ coupled to the quantum field $\phi$ takes the form
\begin{equation}
	j(x)=e\sum_{i=1}^{2}\chi_{i}(t)\,\delta^{(3)}[\mathbf{x}-\mathbf{z}_{i}(t)]
\end{equation}
which says that the internal degree of freedom $\chi_{i}$ of the detector $S_{i}$ is coupled with the environment field $\phi$ at the location $\mathbf{z}_{i}$ of $S_{i}$.

We express the evolution of the density matrix of the whole system (two detectors+the environmental field) using the frameworks of closed-time-path integral formalism.  After carrying out the path integrals of the field, we arrive at a description of the field-influenced two-detector system, governed by the effective action
\begin{align}
	S_{CG}[\mathbf{R},\pmb{\Delta}]&=\int\!ds\;\frac{m}{2}\biggl\{\Bigl[\dot{\mathbf{R}}^{T}(s)\cdot\dot{\pmb{\Delta}}(s)+\dot{\pmb{\Delta}}{}^{T}(s)\cdot\dot{\mathbf{R}}(s)\Bigr]+\Bigl[\mathbf{R}^{T}(s)\cdot\pmb{\Omega}\cdot\pmb{\Delta}(s)+\pmb{\Delta}^{T}(s)\cdot\pmb{\Omega}\cdot\mathbf{R}(s)\Bigr]\biggr\}\notag\\
	&\qquad\qquad+e^{2}\int\!dsds'\;\biggl\{\pmb{\Delta}^{T}(s)\cdot\mathbf{G}_{R}(s,s')\cdot\mathbf{R}(s')+\frac{i}{2}\,\pmb{\Delta}^{T}(s)\cdot\mathbf{G}_{H}(s,s')\cdot\pmb{\Delta}(s')\biggr\}\,,
\end{align}
where the row vector $\pmb{\chi}_{\pm}^{T}=(\chi_{1}^{(\pm)},\chi_{2}^{(\pm)})$ denotes the internal degrees of freedom of the detectors in the $\pm$ branches of the closed time path. The superscript $T$ represents the matrix transposition. Moreover
\begin{align}
	\pmb{\chi}_{\pm}&=\mathbf{R}\pm\frac{1}{2}\,\pmb{\Delta}\,,&\mathbf{\Omega}&=\begin{pmatrix}\omega_{b}^{2}&\sigma\\\sigma &\omega_{b}^{2}\end{pmatrix}\,,&\mathbf{G}(s,s')=\begin{pmatrix}G(\mathbf{z}_{1},s;\mathbf{z}_{1},s')&G(\mathbf{z}_{1},s;\mathbf{z}_{2},s')\\G(\mathbf{z}_{2},s;\mathbf{z}_{1},s')&G(\mathbf{z}_{2},s;\mathbf{z}_{2},s')\end{pmatrix}\,,
\end{align}
with two kinds of correlation functions $G(x;x')$ of the field defined by
\begin{align}
	G_{R}(x;x')&=i\,\theta(t-t')\operatorname{Tr}_{\phi}\Bigl\{\rho_{\beta}\bigl[\phi(x),\phi(x')\bigr]\Bigr\}\,,&G_{H}(x;x')&=\frac{1}{2}\,\operatorname{Tr}_{\phi}\Bigl\{\rho_{\beta}\bigl\{\phi(x),\phi(x')\bigr\}\Bigr\}\,.
\end{align}
The former describes the dissipation effects due to the environment in the evolution of the internal degrees of freedom in the forms of a local self-force and a nonlocal  influence, while the latter summarizes the ramification of quantum fluctuations of the environment upon the internal degree of freedom. The trace is taken over the degrees of freedom of the environment field, and the density matrix in the trace indicates that the initial state of the field is  assumed to be in a thermal state at  temperature $\beta^{-1}$. When the environmental state possesses translation-invariant symmetry, the correlation function $G(x;x')$ has the property $G(x;x')=G(x-x')$ Or, written in explicit space and time variables $G(\mathbf{x},t;\mathbf{x}',t')=G(\mathbf{x}-\mathbf{x}',t-t')$.  The Hadamard function $G_{H}$ is symmetric in its arguments: $G_{H}(x;x')=G_{H}(x';x)$.

We observe that in the effective action $S_{CG}$ we have $\dot{\mathbf{R}}^{T}(s)\cdot\dot{\pmb{\Delta}}(s)=\dot{\pmb{\Delta}}{}^{T}(s)\cdot\dot{\mathbf{R}}(s)$ and $\mathbf{R}^{T}(s)\cdot\pmb{\Omega}\cdot\pmb{\Delta}(s)=\pmb{\Delta}^{T}(s)\cdot\pmb{\Omega}\cdot\mathbf{R}(s)$. Furthermore its imaginary part can be written as a stochastic ensemble average of a Gaussian distribution by the well-known Feynman-Vernon \cite{FeyVer} path integral identity,
\begin{align}
	\exp\biggl[-\frac{1}{2}\int\!dsds'\;\pmb{\Delta}^{T}(s)\cdot\mathbf{G}_{H}(s,s')\cdot\pmb{\Delta}(s')\biggr]&=\int\mathcal{D}\pmb{\xi}\;\mathcal{P}[\pmb{\xi}]\,\exp\biggl[i\int\!ds\;\pmb{\Delta}^{T}(s)\cdot\pmb{\xi}(s)\biggr]\,,
\end{align}
where the stochastic Gaussian distribution $\mathcal{P}[\pmb{\xi}]$ takes the form
\begin{equation}
	\mathcal{P}[\pmb{\xi}]=\mathcal{N}\,\exp\biggl[-\int\!dsds'\;\pmb{\xi}^{T}(s)\cdot\mathbf{G}_{H}^{-1}(s,s')\cdot\pmb{\xi}(s')\biggr]\,,
\end{equation}
and $\mathcal{N}$ is the normalization factor. By doing so, we may rewrite the (complex-valued)  coarse-grained effective action $S_{CG}[\mathbf{R},\pmb{\Delta}]$ \cite{cgea} as a real stochastic effective action~\cite{sea} $S_{eff}[\mathbf{R},\pmb{\Delta},\pmb{\xi}]$, weighted by the distribution $\mathcal{P}[\pmb{\xi}]$. The new variable $\pmb{\xi}$ satisfies the statistics 
\begin{align}
	\langle\pmb{\xi}\rangle&=0\,,&\langle\pmb{\xi}(s)\pmb{\xi}^{T}(s')\rangle&=e^{2}\,\mathbf{G}_{H}(s,s')\,.
\end{align}
Since its distribution is Gaussian, all higher moments can be expanded by the second moment via the Wick's theorem. It can be interpreted as the stochastic driving force on the motion of the internal degrees of freedom of the detectors in their equations of motion obtained by varying $S_{eff}$,\
\begin{align}
	\frac{\delta S_{eff}}{\delta\pmb{\Delta}(s)}&=0\,,&&\Rightarrow&m\,\ddot{\mathbf{R}}(s)+m\,\pmb{\Omega}\cdot\mathbf{R}(s)-e^{2}\int^{s}_{0}\!ds'\;\mathbf{G}_{R}(s,s')\,\mathbf{R}(s')&=\pmb{\xi}(s)\,.
\end{align}
Writing the equation by components may better reveal their physics contents: 
\begin{align}
	m\,\ddot{\chi}_{1}(s)+m\omega_{b}^{2}\,\chi_{1}(s)+m\sigma\,\chi_{2}(s)-e^{2}\int_{0}^{t}\!ds'\;\Bigl[G_{R}(\mathbf{z}_{1},s;\mathbf{z}_{1},s')\chi_{1}(s')+G_{R}(\mathbf{z}_{1},s;\mathbf{z}_{2},s')\chi_{2}(s')\Bigr]&=\xi_{1}(s)\,,\label{E:eqijnk1}\\
	m\,\ddot{\chi}_{2}(s)+m\omega_{b}^{2}\,\chi_{2}(s)+m\sigma\,\chi_{1}(s)-e^{2}\int_{0}^{t}\!ds'\;\Bigl[G_{R}(\mathbf{z}_{2},s;\mathbf{z}_{2},s')\chi_{2}(s')+G_{R}(\mathbf{z}_{2},s;\mathbf{z}_{1},s')\chi_{1}(s')\Bigr]&=\xi_{2}(s)\,.\label{E:eqijnk2}
\end{align}

This set of equations of motion~\eqref{E:eqijnk1} and \eqref{E:eqijnk2} is of great physical interest in that it contains 1) local external harmonic potential, 2) direct interaction between the internal degrees of freedom of the detectors, 3) self-force on each detector  4) stochastic driving force and 5) indirect, retarded interaction between the internal degrees of freedom via the field. The last three effects result from the idf's coupling with the environment, and in particular, the last one where the two detectors idf interact indirectly, as mediated by the environment,  which is totally different in nature from their direct coupling.  This retarded interaction term is weaker as two oscillators are further separated apart.  Note that  if the  oscillators carry a charge and are coupled with a quantized electromagnetic field, the corresponding retardation interaction term also accounts for the effects of the electromagnetic field produced by the nontrivial motions of the other charged oscillators.

\subsubsection{Properties of the Retarded Green's Function}

We first examine the retarded Green's functions $G_{R}(x;x')$ of the scalar quantum field $\phi(x)$, as they describe the nonlocal retarded influence of the environment. 
\begin{align}\label{E:ddfnwiei}
	G_{R}(x;x')&=i\,\theta(t-t')\,\bigl[\phi(x),\phi(x')\bigr]=\frac{1}{4\pi r}\,\theta(\tau)\,\Bigl[\delta(\tau-r)-\delta(\tau+r)\Bigr]\,,
\end{align}
with $\tau=t-t'$ and $r=\lvert\mathbf{x}-\mathbf{x}'\rvert$. We first look into how $G_{R}(x;x')$ propagates between $\chi_i$. {Eq.~\eqref{E:ddfnwiei} implies the nonlocal expressions in, say, \eqref{E:eqijnk1} will have the effects,}
\begin{align}
	\int_{0}^{s}\!ds'\;G_{R}(\mathbf{z}_{1},s;\mathbf{z}_{2},s')\chi_{2}(s')&=\frac{1}{4\pi\ell}\int_{0}^{s}\!ds'\;\delta(s-s'-\ell)\,\chi_{2}(s')=\frac{\theta(s-\ell)}{4\pi\ell}\,\chi_{2}(s-\ell)
\end{align}
with $\ell=\lvert\mathbf{z}_{1}-\mathbf{z}_{2}\rvert$. The theta function ensures causality, namely, starting the motion at $s=0$, then $\chi_{1}$ will not receive the retarded influence from $\chi_{2}$ until at least $s=\ell$.  On the other hand {in the limit $r\to0$}, 
\begin{equation}\label{E:ddfnwiej}
	G_{R}(\mathbf{x},t;\mathbf{x},t')=-\frac{1}{2\pi}\,\theta(\tau)\,\delta'(\tau)\,,
\end{equation}
{and thus \eqref{E:ddfnwiej} in part will} account for a purely local effect, only at the locations of the detectors, 
\begin{align}\label{E:cjhjds}
	\int_{0}^{s}\!ds'\;G_{R}(\mathbf{z}_{1},s;\mathbf{z}_{1},s')\chi_{1}(s')&=\frac{1}{2\pi}\left[\delta(s-s')\,\chi_{1}(s')\bigg|_{s'=0}^{s'=s}-\int_{0}^{s}\!ds'\;\delta(s-s')\,\chi'_{1}(s')\right]\,,
\end{align}
{due to the reaction of the radiation field}. The first term on the righthand side is not well-defined unless we introduce a energy cutoff for the environment, whereby the resulting cutoff-dependent expression can be absorbed as a frequency normalization of $\chi_{1}$.   In other words, the natural frequency of $\chi$ undergoes a  shift due to its interaction with the environment.  Now \eqref{E:cjhjds} becomes
\begin{equation}
	\int_{0}^{s}\!ds'\;G_{R}(\mathbf{z}_{1},s;\mathbf{z}_{1},s')\chi_{1}(s')=-\frac{m\,\delta\omega^{2}}{e^{2}}\,\chi_{1}(s)-\frac{1}{4\pi}\,\dot{\chi}_{1}(s)\,.
\end{equation} 
Putting  them back into the equations of motion, we arrive at
\begin{align}
	\ddot{\chi}_{1}(s)+2\gamma\,\dot{\chi}_{1}(s)-2\gamma\,\frac{\theta(s-\ell)}{\ell}\,\chi_{2}(s-\ell)+\omega^{2}\,\chi_{1}(s)+\sigma\,\chi_{2}(s)&=\frac{1}{m}\,\xi_{1}(s)\,,\\
	\ddot{\chi}_{2}(s)+2\gamma\,\dot{\chi}_{2}(s)-2\gamma\,\frac{\theta(s-\ell)}{\ell}\,\chi_{1}(s-\ell)+\omega^{2}\,\chi_{2}(s)+\sigma\,\chi_{1}(s)&=\frac{1}{m}\,\xi_{2}(s)\,,
\end{align}
where $\gamma=e^{2}/8\pi m$ is the damping constant, and $\omega^{2}=\omega_{b}^{2}+\delta\omega^{2}$ is the renormalized natural frequency.   Formally this frequency shift $\delta\omega^{2}$ is written as $\delta\omega^{2}=-4\gamma\,\delta(0)$, but we can put it in a form containing the cutoff scale $\Lambda$ of the model in consideration,
\begin{equation}
	\delta\omega^{2}=-4\gamma\,\delta(0)=-\frac{2\gamma}{\pi}\int_{-\infty}^{\infty}\!d\kappa=-\frac{4\gamma}{\pi}\int_{0}^{\Lambda}\!d\kappa=-\frac{4\gamma\Lambda}{\pi}\,.
\end{equation}

\subsubsection{Idf Variables $\chi_\pm$:  Symmetric (Sum or Center of Mass) and Antisymmetric (Difference or Relative) Normal Modes}

We can decouple this set of equations by superpositions $\chi_{+}=(\chi_{1}+\chi_{2})/2$ and $\chi_{-}=\chi_{1}-\chi_{2}$,
\begin{align}
	\ddot{\chi}_{+}+2\gamma\,\dot{\chi}_{+}(s)-2\gamma\,\frac{\theta(s-\ell)}{\ell}\,\chi_{+}(s-\ell)+\omega^{2}_{+}\,\chi_{+}(s)&=\frac{1}{m}\,\xi_{+}(s)\,,\label{E:iuwrakjd1}\\
	\ddot{\chi}_{-}+2\gamma\,\dot{\chi}_{-}(s)+2\gamma\,\frac{\theta(s-\ell)}{\ell}\,\chi_{-}(s-\ell)+\omega^{2}_{-}\,\chi_{-}(s)&=\frac{1}{m}\,\xi_{-}(s)\,,\label{E:iuwrakjd2}
\end{align}
with $\xi_{+}=(\xi_{1}+\xi_{2})/2$, $\xi_{-}=\xi_{1}-\xi_{2}$ and $\omega_{\pm}^{2}=\omega^{2}\pm\sigma$. In the {small-separation} limit $\omega_{\pm}\ell\ll1$, we Taylor-expand the retarded expressions into $\chi_{\pm}(s-\ell)\simeq\chi_{\pm}(s)-\ell\,\dot{\chi}_{\pm}(\tau)+\cdots$ and observe that
\begin{align}
	&\quad\;\ddot{\chi}_{\pm}(s)+2\gamma\,\dot{\chi}_{\pm}(s)+\omega_{\pm}^{2}\chi_{\pm}(s)\mp2\gamma\,\frac{\theta(s-\ell)}{\ell}\,\chi_{\pm}(s-\ell)\notag\\
	&\simeq\ddot{\chi}_{\pm}(s)+2\gamma\,\dot{\chi}_{\pm}(s)+\bigl(\omega_{b\,\pm}^{2}+\delta\omega^{2}\bigr)\chi_{\pm}(s)\mp\frac{2\gamma}{\ell}\Bigl[\chi_{\pm}(s)-\ell\,\dot{\chi}_{\pm}(s)+\frac{\ell^{2}}{2}\,\ddot{\chi}_{\pm}(s)-\frac{\ell^{3}}{3!}\,\dddot{\chi}_{\pm}(s)+\cdots\Bigr]\notag\\
	&\simeq\bigl(1\mp\gamma\ell\bigr)\,\ddot{\chi}_{\pm}(s)+2\gamma\bigl(1\pm1\bigr)\,\dot{\chi}_{\pm}(s)+\varpi^{2}_{\pm}\chi_{\pm}(s)\pm\frac{\gamma\ell^{2}}{3}\,\dddot{\chi}_{\pm}(s)+\cdots\,,\label{E:qarens}
\end{align}
with $\varpi_{\pm}^{2}=\omega_{b\pm}^{2}-\dfrac{4\gamma\Lambda}{\pi}\mp\dfrac{2\gamma}{\ell}$ and $\omega_{b\pm}^{2}=\omega_{b}^{2}\pm\sigma$.  

\subsubsection{Severely Restricted Validity Range of Zero-Separation Results}

In choosing $\Lambda\ell=\pi/2$, we obtain the equations of motion found in another often used configuration for the investigation of the thermal entanglement in a common bath \cite{Goan,Paz} 
\begin{align}
	\ddot{\chi}_{+}+4\gamma\,\dot{\chi}_{+}(s)+\widetilde{\omega}^{2}_{+}\,\chi_{+}(s)&=\frac{1}{m}\,\xi_{+}(s)\,,\label{E:nckrwq1}\\
	\ddot{\chi}_{-}+\widetilde{\omega}^{2}_{-}\,\chi_{-}(s)&=\frac{1}{m}\,\xi_{-}(s)\,,\label{E:nckrwq2}
\end{align}
where $\widetilde{\omega}_{-}^{2}=\omega_{b\,-}^{2}$ but $\widetilde{\omega}_{+}^{2}=\omega_{b\,+}^{2}-8\gamma\Lambda/\pi$. In that configuration, two coupled oscillators are assumed to be placed at the same spatial location and interact with a common thermal bath. With the retardation effects completely ignored, the relative mode becomes a undamped driven oscillator.  This allows for the initial information of the system to persist and its effects are oscillatory in time. This feature fundamentally changes the characteristics of the late time entanglement in a common bath, in comparison with the finite separation case we are studying here.  

We can derive the validity range of prior results of  the same-location configuration. Keeping the third-order time derivative in the small-distance expansion {\eqref{E:qarens}}, we find the equation of motion for the relative mode becomes
\begin{equation}
	(1+\gamma\ell\bigr)\ddot{\chi}_{-}(s)-\frac{\gamma\ell^{2}}{3}\,\dddot{\chi}_{-}(s)+\bigl(\omega_{-}^{2}+\frac{2\gamma}{\ell}\bigr)\chi_{\pm}(s)+\cdots=\frac{1}{m}\,\xi_{-}(s)\,.
\end{equation}
The third-order time derivative term now plays the role of a friction force.  Compared with the corresponding equation of motion for the CoM mode,  the damping of the relative mode is much weaker by an order of $(\omega\ell)^{2}$.  Based on \eqref{E:nckrwq1} and \eqref{E:nckrwq2} we can say that the results for the configuration of  two oscillators in the same location interacting with a common bath  can at best  be valid in the very short time before the dynamics of the relative mode is fully relaxed, that is, within the time scale much shorter than $(\gamma\omega_{-}^{2}\ell^{2})^{-1}$.

\subsection{Solutions and Dynamics}

We now investigate the dynamics of the system derived from the solutions of the equations of motion~\eqref{E:eqijnk1} and \eqref{E:eqijnk2}.

\subsubsection{An iterative scheme to obtain early time solutions}

As a brief interlude, we first mention a method to obtain early time solutions. {This gives  a clear depiction of how mutual influences are transmitted back and forth between the internal degrees of freedom of both detectors.}

The equations of motion of the normal modes are second-order differential equations with fixed delay. One way to solve the equations of this type is to make use of iteration. We divide the evolution time $t$ into intervals of length $\ell$, that is, $t\in[(n-1)\ell,n\ell]$ with $n\in\mathbb{N}$. When $t$ falls in the first interval $t\in[0,\ell]$, the delayed term in~\eqref{E:eqijnk1} and \eqref{E:eqijnk2} has no effect, so the solutions to \eqref{E:eqijnk1} and \eqref{E:eqijnk2} can be found exactly. When time evolves to the next interval $t\in[\ell, 2\ell]$, the delayed term has to be taken into consideration, but since $t-\ell\in[0,\ell]$, we can substitute the previously found solution to the delay term. In so doing the equations of motion in the time $t\in[\ell, 2\ell]$ become ordinary differential equations and they are exactly solvable. We may proceed with the same procedures to the next time interval $t\in[2\ell, 3\ell]$ and further on. This technique is straightforward but becomes very cumbersome after a few iterations, Besides, in this iteration scheme, it is not straightforward to identify the existence of the relaxed motion, so it is merely suitable for finding the early-time solutions to~\eqref{E:eqijnk1} and \eqref{E:eqijnk2}. Since we are interested in the late time behaviors of the solutions, we will not pursue this approach any further. 

\subsubsection{Systematic method for seeking the solutions}

To systematically find the solutions to the equations of motion~\eqref{E:eqijnk1} and \eqref{E:eqijnk2}, we may perform the Laplace transformation on the equations of motion
\begin{align}
	&&&\Bigl[-\dot{\chi}_{\pm}(0)-s\,\chi_{\pm}(0)+s^{2}\,\widetilde{\chi}_{\pm}(s)\Bigr]+2\gamma\,\Bigl[-\chi_{\pm}(0)+s\,\widetilde{\chi}_{\pm}(s)\Bigr]+\omega^{2}_{\pm}\,\widetilde{\chi}_{\pm}(s)\mp\frac{2\gamma}{\ell}\,e^{-sd}\,\widetilde{\chi}_{\pm}(s)=\frac{1}{m}\,\widetilde{\xi}_{\pm}(s)\,,\notag\\
	&\Rightarrow&&\Bigl[s^{2}+2\gamma s+\omega_{\pm}^{2}\mp\frac{2\gamma}{\ell}\,e^{-s\ell}\Bigr]\,\widetilde{\chi}_{\pm}(s)=\frac{1}{m}\,\widetilde{\xi}_{\pm}(s)+\Bigl[\dot{\chi}_{\pm}(0)+\bigl(s+2\gamma\bigr)\,\chi_{\pm}(0)\Bigr]\,,\label{E:bkuewja}
\end{align}
where we have used the following identities
\begin{align}
	\int_{0}^{\infty}\!dt\;\dot{\chi}(t)\,e^{-st}&=-\chi(0)+s\,\widetilde{\chi}(s)\,,\\
	\int_{0}^{\infty}\!dt\;\ddot{\chi}(t)\,e^{-st}&=-\dot{\chi}(0)-s\,\chi(0)+s^{2}\,\widetilde{\chi}(s)\,,\\
	\int_{0}^{\infty}\!dt\;\theta(t-\ell)\,\chi(t-\ell)\,e^{-st}&=e^{-s\ell}\,\widetilde{\chi}(s)\,,
\end{align}
and the Laplace transformation of the function $\chi(t)$ is defined for $t\geq0$ by
\begin{equation}
	\widetilde{\chi}(s)=\int_{0}^{\infty}\!dt\;\chi(t)\,e^{-st}.\,,
\end{equation}
Solving \eqref{E:bkuewja} for $\widetilde{\chi}_{\pm}(s)$ yields
\begin{align}\label{E:xksjrweq} 
	\widetilde{\chi}_{\pm}(s)=\widetilde{d}_{1}^{(\pm)}(s)\,\chi_{\pm}(0)+\widetilde{d}_{2}^{(\pm)}(s)\,\dot{\chi}_{\pm}(0)+\frac{1}{m}\,\widetilde{d}_{2}^{(\pm)}(s)\,\widetilde{\xi}_{\pm}(s)\,,
\end{align}
where 
\begin{align}\label{E:qjskerw}
	\widetilde{d}_{1}^{(\pm)}(s)&=\frac{s+2\gamma}{\widetilde{g}_{\pm}(s)}\,,&\widetilde{d}_{2}^{(\pm)}(s)&=\frac{1}{\widetilde{g}_{\pm}(s)}\,,
\end{align}
and $\displaystyle\widetilde{g}_{\pm}(s)=s^{2}+2\gamma s+\omega_{\pm}^{2}\mp\frac{2\gamma}{\ell}\,e^{-s\ell}$. From the dependence on the initial conditions, we identify that the first two terms on the righthand side of \eqref{E:xksjrweq} correspond to the homogenous solutions to the equation of motion while the third term correspond to the inhomogeneous solution.  We can find the time evolution of $\chi_{\pm}(t)$ by performing the inverse Laplace transformation on \eqref{E:xksjrweq},
\begin{equation}\label{E:neisnkfjw}
	\chi_{\pm}(t)=\frac{1}{2\pi i}\int_{\Gamma}\!ds\;\widetilde{\chi}_{\pm}(s)\,e^{st}\,,
\end{equation}
where the Bromwich contour $\Gamma$ is so chosen to make the integral in \eqref{E:neisnkfjw} well-defined.

We use the residue theorem to evaluate the contour integral in \eqref{E:neisnkfjw}. All we need to do is to identify the poles associated with $\widetilde{d}_{1,2}^{(\pm)}(s)$. This is equivalent to finding the solutions to $\widetilde{g}_{\pm}(s)=0$. Compared with the corresponding equations in the private bath case, the function $\widetilde{g}_{\pm}(s)$ has an extra term proportional to $e^{-s\ell}$ and this renders $\widetilde{g}_{\pm}(s)$ a transcendental function, whose exact analytical solution is typically hard to come by. Some approximation schemes are needed.  \\

\paragraph*{1. Strong Damping Case}  When $2\gamma\ell\gg1$ and $\gamma>\omega$, the function $\widetilde{g}_{\pm}(s)=0$ reduces to
\begin{align}
	y^{2}+2\gamma\ell y+\omega_{\pm}^{2}\ell^{2}\mp2\gamma\ell\,e^{-y}&=0&&\longrightarrow&2\gamma\ell y+\omega_{\pm}^{2}\ell^{2}\mp2\gamma\ell\,e^{-y}&=0\,,
\end{align}
because $\lvert y\rvert$ must be smaller than $2\gamma\ell$ otherwise $\widetilde{g}_{\pm}(s)$ will be too positive to be equal to zero. Solving 
$$2\gamma\ell y+\omega_{\pm}^{2}\ell^{2}\mp2\gamma\ell\,e^{-y}=0$$ gives
\begin{align}\label{E:woijeow}
	y=-\frac{\omega_{\pm}^{2}\ell}{2\gamma}+\Psi(\pm\,e^{\frac{\omega_{\pm}^{2}\ell}{2\gamma}})\,,
\end{align}
where $\Psi(z)$ is the principal solution for $x$ in $z=x\,e^{x}$. If we look for a real solution then the relative mode does not have any, because $\Psi(z)$ is a complex function of $z$ when $z<0$. On the other hand since $\Psi(z)$ is real for $z\geq0$, it is straightforward to see that $y$ becomes positive when $2\gamma>\omega_{\pm}^{2}\ell$. A positive real solution to $\widetilde{g}(s)_{\pm}=0$ denotes a runaway solution when we apply the residue theorem to the inverse Laplace transformation like \eqref{E:neisnkfjw}. The culprit for the existence of the runaway solution can be seen explicitly to be the delayed term in the equations of motion~\eqref{E:eqijnk1} and \eqref{E:eqijnk2}. Thus we can highlight a stark contrast between the Markovian and the non-Markovian motion. \\

\paragraph*{Markovian dynamics}  For the Markovian motion of the oscillator, the function $\widetilde{g}_{\pm}(s)$ typically takes the form
\begin{align}
	\widetilde{g}_{\pm}(s)&=s^{2}+2\gamma s+\omega_{\pm}^{2}\,,&&\text{the Markovian case}\,,
\end{align}
so the strong damping $\gamma>\omega_{\pm}$ merely causes overdamped motion of the oscillator since in this case $\widetilde{g}_{\pm}(s)=0$ gives
\begin{align}
	s=-\,\gamma\pm\sqrt{\gamma^{2}-\omega_{\pm}^{2}}<0\,.
\end{align} \\

\paragraph*{non-Markovian dynamics}

However, in the non-Markovian case as shown previously, strong damping can imply a real and positive $s$ and in turn unstable motion. In principle, a large damping constant $\gamma$ is supposed to efficiently damp out  the oscillator's motion. On the other hand, it also means that the oscillator interacts strongly with the environment. Thus when two oscillators couple with a common environment, their mutual influences will last and the exchange between one another goes on much longer in  the course of time evolution. On top of that, when $2\gamma>\omega_{\pm}^{2}\ell$, each reciprocal mutual influence is enhanced by a factor $\dfrac{2\gamma}{\omega_{\pm}^{2}\ell}>1$ and is accumulated in such a way that it adds up constructively, especially for the CoM mode. This acts counter to the damping due to the dissipative force and finally wins to become a runaway solution.  In contrast, the Markovian motion does not have the retarded terms, so the corresponding motion just rapidly damps away in the strong damping case. \\

\paragraph*{ 2.   Weak Damping Case} 

We first locate the solutions to $\widetilde{g}_{\pm}(s)=0$ for large $\lvert s\rvert\ell\gg\omega_{\pm}\ell$, and write $\widetilde{g}_{\pm}(s)$ as
\begin{align}\label{E:cmsfs}
	s^{2}e^{s\ell}\Bigl[1+\frac{2\gamma}{s}+\frac{\omega_{\pm}^{2}}{s^{2}}\Bigr]&=\pm\frac{2\gamma}{\ell}\,,&&\Rightarrow&y^{2}e^{y}&\simeq\pm2\gamma\ell=2\gamma\ell\,e^{i\,(2n+\frac{1}{2}\mp\frac{1}{2})\pi}\,,
\end{align}
with $y=s\ell$ and $n\in\mathbb{Z}$.   Taking the logarithm of both sides of \eqref{E:cmsfs} gives
\begin{equation}\label{E:xknskks}
	y+2\ln y\simeq\ln2\gamma\ell+i\,\pi\bigl(2n+\frac{1}{2}\mp\frac{1}{2}\bigr)\,.
\end{equation}
Now substitute $y=u+i\,v=\sqrt{u^{2}+v^{2}}\,e^{i\,\theta}$ with $u$, $v$, $\theta\in\mathbb{R}$ into \eqref{E:xknskks}, and we obtain
\begin{align}\label{E:ajnfkwe}
	\Bigl[u+\ln\bigl(u^{2}+v^{2}\bigr)-\ln2\gamma\ell\Bigr]+i\Bigl[v+2\tan^{-1}\frac{v}{u}-\pi\bigl(2n+\frac{1}{2}\mp\frac{1}{2}\bigr)\Bigr]=0\,.
\end{align}
To fulfill \eqref{E:ajnfkwe}, the imaginary part of \eqref{E:ajnfkwe} implies $\lvert v\rvert\gg\lvert u\rvert$ such that 
\begin{align}
	\ln\bigl(u^{2}+v^{2}\bigr)&\simeq2\ln\lvert v\rvert\,,&\tan^{-1}\frac{v}{u}&\simeq\operatorname{sgn}(v)\,\frac{\pi}{2}\,,
\end{align}
and then it gives $v$,
\begin{align}
	v&\simeq\pi\,\Bigl[2n+\frac{1}{2}\mp\frac{1}{2}-\operatorname{sgn}(v)\Bigr]\,,&&\text{with}&n&\gg1\,.
\end{align}
If we put the solved $v$ back into the real part of  \eqref{E:ajnfkwe}, we obtain
\begin{equation}\label{E:jwehaaape}
	u\simeq\ln2\gamma\ell-2\ln\lvert v\rvert=\ln2\gamma\ell-2\ln\left\{\Bigl[2n+\frac{1}{2}\mp\frac{1}{2}-\operatorname{sgn}(v)\Bigr]\,\pi\right\}\,.
\end{equation}
Unless $\gamma\ell\gg1$, typically $u<0$ for large $n$. Since the negative values of $u$ will contribute to the decaying behavior of \eqref{E:qjskerw} once we apply the residue theorem to \eqref{E:neisnkfjw},  the factor $e^{\operatorname{Re}\{s\}t}$ behaves like
\begin{align}
	\exp\biggl[\frac{u}{\ell}\,t\biggr]\simeq\exp\biggl\{-\frac{2t}{\ell}\ln\frac{\Bigl[2n+\frac{1}{2}\mp\frac{1}{2}-\operatorname{sgn}(v)\Bigr]\,\pi}{\sqrt{2\gamma\ell}}\biggr\}=\mathcal{O}(\frac{1}{n^{2t/\ell}})\,.
\end{align}
Therefore at late times $t\gg\ell$, the contributions to the solution \eqref{E:neisnkfjw} from the large $n$ poles are heavily suppressed. Alternative, we may compare $\lvert u\rvert/\ell$ with the damping constant $\gamma$. From the argument that follows \eqref{E:jwehaaape}, we see that we typically have $\lvert u\rvert\gg\gamma\ell$, in particular for the case of weak damping $\gamma/\omega\ll1$. We conclude that the factor $e^{ut}$ will fall off much faster than $e^{-\gamma t}$ for large $n$ at late time.

When both oscillators are far away from one another, the delayed term in the equations of motion~\eqref{E:eqijnk1} and \eqref{E:eqijnk2} is suppressed by the factor $\gamma/\ell$, so we may treat this delayed term as a small perturbation. Note that in this case since the non-Markovian or the retardation effects are very feeble, the configuration and the equations of motion are almost identical to those in the case of the private baths of the same temperature. However, the similarities are merely superficial because there is no environment correlation in the private bath case.\\

\paragraph*{3.  Late time behavior}

At this point we see that the \textit{dominant late-time contribution }to \eqref{E:neisnkfjw} should come from the pole with the smallest value of $\lvert u\rvert$. We first consider the case $\gamma\ll\omega_{\pm}^{2}\ell$, which covers the limits of weak damping and/or large distance between oscillators. The delayed term will be treated as a small perturbation. Let the function $\widetilde{g}_{\pm}(s)$ be written as 
\begin{align}\label{E:kwkwkdjfs}
	\widetilde{g}_{\pm}(s)=\frac{y^{2}}{\ell^{2}}+\omega_{\pm}^{2}+\frac{2\gamma}{\ell}\Bigl[y\mp e^{-y}\Bigr]\,,
\end{align}
with $y=s\ell$. The zeros of $\widetilde{g}_{\pm}(s)$ to the lowest order in $\gamma/\ell$ is simply given by $i\,\omega_{\pm}\ell$ and the minus of them. Here we only demonstrate the $i\,\omega_{\pm}\ell$ case, and the result for the minus sign case can be found accordingly. Next if we take into account the contribution of the order $\gamma/\ell$ in \eqref{E:kwkwkdjfs}, we should expect the correction to the solution is also at least of the order $\gamma/\ell$.  Writing  $y=i\,\omega_{\pm}\ell+\dfrac{\gamma}{\ell}\,(\alpha+i\,\beta)+\mathcal{O}(\dfrac{\gamma}{\ell})^{2}$ with $\alpha$, $\beta\in\mathbb{R}$, and substituting this into \eqref{E:kwkwkdjfs} we can find the corrections to the solutions due to the inclusion of the terms of the order $\mathcal{O}(\dfrac{\gamma}{\ell})$,  
\begin{align}
	\alpha&=-\frac{\ell}{\omega_{\pm}}\Bigl[\omega_{\pm}\ell\pm\sin\omega_{\pm}\ell\Bigr]\,,&\beta&=\mp\frac{\ell}{\omega_{\pm}}\,\cos\omega_{\pm}\ell\,.
\end{align}
Thus the fast mode that the zeros of $\widetilde{g}_{+}(s)$ are given by
\begin{equation}
	s_{+}=\pm\,i\,\omega_{+}+\gamma\Bigl[-\Bigl(1+\frac{\sin\omega_{+}\ell}{\omega_{+}\ell}\Bigr)\mp i\,\frac{\cos\omega_{+}\ell}{\omega_{+}\ell}\Bigr]=\pm\, i\,\omega_{+}\left[1-\frac{\gamma}{\omega_{+}}\frac{\cos\omega_{+}\ell}{\omega_{+}\ell}\right]-\gamma\left[1+\frac{\sin\omega_{+}\ell}{\omega_{+}\ell}\right]\,,
\end{equation}
up to the order $\mathcal{O}(\dfrac{\gamma}{\ell})^{2}$. Likewise for the slow mode we obtain the zeros of $\widetilde{g}_{-}(s)$ with
\begin{equation}
	s_{-}=\pm\,i\,\omega_{-}+\gamma\Bigl[-\Bigl(1-\frac{\sin\omega_{-}\ell}{\omega_{-}\ell}\Bigr)\pm i\,\frac{\cos\omega_{-}\ell}{\omega_{-}\ell}\Bigr]=\pm\, i\,\omega_{-}\left[1+\frac{\gamma}{\omega_{-}}\frac{\cos\omega_{-}\ell}{\omega_{-}\ell}\right]-\gamma\left[1-\frac{\sin\omega_{-}\ell}{\omega_{-}\ell}\right]\,.
\end{equation}
We observe that the contribution from the retarded term is smaller with large separation $\ell$ between oscillators, which is consistent with our physical intuition. We may group the expressions in $s_{\pm}$ in such a way to define the effective damping $\Gamma_{\pm}$ and the effective oscillating frequency $W_{\pm}$,
\begin{align}\label{E:oerndj}
	\Gamma_{\pm}&=\gamma\left[1\pm\frac{\sin\omega_{\pm}\ell}{\omega_{\pm}\ell}\right]\,,&W_{\pm}&=\omega_{\pm}\left[1\mp\frac{\gamma}{\omega_{\pm}}\frac{\cos\omega_{\pm}\ell}{\omega_{\pm}\ell}\right]\,.
\end{align}
The effective damping constant is always nonnegative and can  only for the relative mode in the limit $\omega_{-}\ell\to0$; however such a small separation can induce instability in oscillators' motion because $W_{\pm}^{2}$ 
\begin{equation}\label{E:nvkjske}
	W_{\pm}^{2}=\omega_{\pm}^{2}\left[1\mp\frac{2\gamma}{\omega_{\pm}}\frac{\cos\omega_{\pm}\ell}{\omega_{\pm}\ell}\right]\,,
\end{equation} 
becomes negative when $2\gamma>\omega_{\pm}^{2}\ell$. Negative $W^{2}_{\pm}$ implies a runaway solution. In addition the condition $2\gamma>\omega_{\pm}^{2}\ell$ also hints at \textit{the possibility that the non-Markovian motion tends to be unstable for strong damping}. Therefore in the following discussion, we will restrict ourselves to the condition $2\gamma<\omega_{\pm}^{2}\ell$ so that we have a well-defined effective oscillating frequency. Finally we remark that this condition is relatively more stringent for the relative mode because when the inter-oscillator coupling $\sigma$ is strong, the relative mode frequency $\omega_{-}$ tends to be small and it makes the condition harder to meet. Furthermore, the perturbative descriptions in \eqref{E:oerndj}, though simple in form, tend to generate spurious wiggling with $\ell$, due to the sinusoidal expression in \eqref{E:nvkjske}. This wiggling will diminish with large values of $\ell$.

\subsubsection{A third method}

Yet another useful approach in dealing with the \textit{non-Markovian dynamics} is to directly Taylor-expand $1/\widetilde{g}_{\pm}(s)$ in terms of the small parameter $\dfrac{2\gamma}{\omega_{\pm}\ell}$ and evaluate the corresponding integrals, instead of resorting to the residue theorem. {This is possible only when we  assume weak coupling with $\dfrac{2\gamma}{\omega_{\pm}^{2}\ell}<1$.} The idea is to write $1/\widetilde{g}_{\pm}(s)$ as
\begin{align}\label{E:ddljnsd}
	\frac{1}{\widetilde{g}_{\pm}(s)}=\frac{1}{s^{2}+2\gamma s+\omega_{\pm}^{2}\mp\dfrac{2\gamma}{\ell}\,e^{-s\ell}}&=\frac{1}{s^{2}+2\gamma s+\omega_{\pm}^{2}}\sum_{n=0}^{\infty}\biggl[\frac{\pm\dfrac{2\gamma}{\ell}\,e^{-s\ell}}{s^{2}+2\gamma s+\omega_{\pm}^{2}}\biggr]^{n}\,.
\end{align}
Owing to the fact
\begin{align}
	\frac{1}{(s^{2}+2\gamma s+\omega_{\pm}^{2})^{n+1}}=\left(-\frac{1}{2}\right)^{n}\frac{1}{n!}\left[\frac{1}{\omega_{\pm}}\frac{\partial}{\partial\omega_{\pm}}\right]^{n}\frac{1}{s^{2}+2\gamma s+\omega_{\pm}^{2}}\,,
\end{align}
we find \eqref{E:ddljnsd} will take the form
\begin{equation}
	\frac{1}{\widetilde{g}_{\pm}(s)}=\sum_{n=0}^{\infty}\left(\mp\frac{\gamma}{\ell}\right)^{n}\frac{e^{-nsl}}{n!}\left[\frac{1}{\omega_{\pm}}\frac{\partial}{\partial\omega_{\pm}}\right]^{n}\frac{1}{s^{2}+2\gamma s+\omega_{\pm}^{2}}\,.
\end{equation}
The calculations based on this expansion seems free from the artifact wiggling. This approach in a sense is similar but more general than the perturbative approach because the perturbative approach can be viewed as a convenient case that $s$ in $e^{-sl}$ is replace by $-i\,\omega_{\pm}$.

For \textsl{weak bath-oscillator interaction}, the \textsl{mutual non-Markovian influence between oscillators }decays quickly after a couple of exchanges, so both approaches will work nicely for the late-time dynamics. This is also the reason why at late time the non-Markovian dynamics can be equivalently summarized into a seemingly Markovian form with the\textit{ effective damping} and the \textit{effective frequency} given by \eqref{E:oerndj}. However, we would like to stress that the bath in the common bath case is correlated at the locations of the oscillators, while the baths in the private bath case are not. This is reflected in the separation dependence of $\widetilde{d}_{1,2}^{(\pm)}(s)$ in the common bath case by the perturbative approach. 


\section{Covariance Matrix}

From \eqref{E:xksjrweq}, the general solutions for the motions of the CoM and the relative modes are described by
\begin{equation}\label{E:nuhiwhd} 
	\chi_{\pm}(t)=d_{1}^{(\pm)}(t)\,\chi_{\pm}(0)+d_{2}^{(\pm)}(t)\,\dot{\chi}_{\pm}(0)+\frac{1}{m}\int_{0}^{t}\!dt'\;d_{2}^{(\pm)}(t-t')\,\xi_{\pm}(t')\,,
\end{equation}
with $d_{1}^{(\pm)}(0)=1$, $\dot{d}_{1}^{(\pm)}(0)=0$, $d_{1}^{(\pm)}(0)=0$, $\dot{d}_{2}^{(\pm)}(0)=1$, and they are all equal to zero for $t<0$. Since we only consider the case $2\gamma/\omega_{\pm}^{2}\ell<1$, we have seen that $d_{2}^{(\pm)}(\tau)$ exponentially decays with $t$. At late time after the motions of the normal modes are fully relaxed, Eq.~\eqref{E:nuhiwhd} asymptotically approaches 
\begin{equation}\label{E:zuhiwhd} 
	\chi_{\pm}(t)=\frac{1}{m}\int_{0}^{t}\!dt'\;d_{2}^{(\pm)}(t-t')\,\xi_{\pm}(t')\,,
\end{equation}
so the initial information contained in $\chi_{\pm}(0)$ and $\dot{\chi}_{\pm}(0)$ will not survive at late time. {Eq.~\eqref{E:zuhiwhd} enables us to compute the late-time values of the covariance matrix elements}.  Before proceeding, we first express the force-force correlations {in terms of the Hadamard functions of the environment},
\begin{align*}
	\langle\xi_{+}(t')\xi_{+}(t'')\rangle&=\frac{1}{4}\,\langle\Bigl[\xi_{1}(t')+\xi_{2}(t')\Bigr]\Bigl[\xi_{1}(t'')+\xi_{2}(t'')\Bigr]\rangle=\frac{e^{2}}{2}\Bigl[G_{H}(\mathbf{0},t'-t'')+G_{H}(\mathbf{z}_{1}-\mathbf{z}_{2},t'-t'')\Bigr]\,,\\
	\langle\xi_{-}(t')\xi_{-}(t'')\rangle&=\langle\Bigl[\xi_{1}(t')-\xi_{2}(t')\Bigr]\Bigl[\xi_{1}(t'')-\xi_{2}(t'')\Bigr]\rangle=2e^{2}\Bigl[G_{H}(\mathbf{0},t'-t'')-G_{H}(\mathbf{z}_{1}-\mathbf{z}_{2},t'-t'')\Bigr]\,,\\
	\langle\xi_{+}(t')\xi_{-}(t'')\rangle&=\frac{1}{2}\,\langle\Bigl[\xi_{1}(t')+\xi_{2}(t')\Bigr]\Bigl[\xi_{1}(t'')-\xi_{2}(t'')\Bigr]\rangle=0\,,
\end{align*}
where we have made use of the facts that $\xi_{+}=(\xi_{1}+\xi_{2})/2$, $\xi_{+}=\xi_{1}-\xi_{2}$ and
\begin{align}
	\langle\xi_{1}(t')\xi_{1}(t'')\rangle&=e^{2}G_{H}(\mathbf{z}_{1},t';\mathbf{z}_{1},t'')=e^{2}G_{H}(\mathbf{0},t'-t'')\,,\\
	\langle\xi_{2}(t')\xi_{2}(t'')\rangle&=e^{2}G_{H}(\mathbf{z}_{2},t';\mathbf{z}_{2},t'')=e^{2}G_{H}(\mathbf{0},t'-t'')\,,\\
	\langle\xi_{1}(t')\xi_{2}(t'')\rangle=\langle\xi_{2}(t')\xi_{1}(t'')\rangle&=e^{2}G_{H}(\mathbf{z}_{1},t';\mathbf{z}_{2},t'')=e^{2}G_{H}(\mathbf{z}_{1}-\mathbf{z}_{2},t'-t'')\,.
\end{align}
Thus we are ready to set up the building blocks to construct the elements of the covariance matrix. {All we need are $\langle\chi_{\pm}^{2}(t)\rangle$, $\dfrac{1}{2}\,\langle\{\chi_{+}(t),\chi_{-}(t)\}\rangle$ and the counterparts for the momentum variables.} We begin with $\langle\chi_{\pm}^{2}(t)\rangle$ and $\dfrac{1}{2}\,\langle\{\chi_{+}(t),\chi_{-}(t)\}\rangle$.

\begin{itemize}
\item {$\langle\chi_{+}^{2}(t)\rangle$}: We rewrite $\chi_{+}(t)$ in \eqref{E:zuhiwhd} in terms $\xi_{1}$ and $\xi_{2}$ and  arrive at {its late-time value}
\begin{align}
	\langle\chi_{+}^{2}(t)\rangle&=\frac{1}{m^{2}}\int_{0}^{t}\!dt'\,dt''\;d_{2}^{(+)}(t-t')d_{2}^{(+)}(t-t'')\,\langle\xi_{+}(t')\xi_{+}(t'')\rangle\notag\\
	&=\frac{e^{2}}{2m^{2}}\int_{0}^{t}\!dt'\,dt''\;d_{2}^{(+)}(t')d_{2}^{(+)}(t'')\Bigl[G_{H}(\mathbf{0},t'-t'')+G_{H}(\mathbf{z}_{1}-\mathbf{z}_{2},t'-t'')\Bigr]\,.
\end{align}
In the limit $t\to\infty$ we arrive at
\begin{equation}
	\lim_{t\to\infty}\langle\chi_{+}^{2}(t)\rangle=\frac{e^{2}}{2m^{2}}\int_{-\infty}^{\infty}\!\frac{d\kappa}{2\pi}\;\lvert\overline{d}_{2}^{(+)}(\kappa)\rvert^{2}\Bigl[\overline{G}_{H}(\mathbf{0},\kappa)+\overline{G}_{H}(\mathbf{z}_{1}-\mathbf{z}_{2},\kappa)\Bigr]\,,\label{E:twiehs1}
\end{equation}
where $\overline{d}_{2}^{(+)}(\kappa)$, $\overline{G}_{H}(\mathbf{0},\kappa)$ are the Fourier transforms of $d_{2}^{(+)}(t'-t'')$, $G_{H}(\mathbf{0},t'-t'')$, for example. We define the Fourier transformation of a function $f(\tau)$ by
\begin{align}
	\overline{f}(\kappa)&=\int_{-\infty}^{\infty}\!d\tau\;f(\tau)\,e^{i\kappa\tau}\,,&f(\tau)&=\int_{-\infty}^{\infty}\!\frac{d\kappa}{2\pi}\;\overline{f}(\kappa)\,e^{-i\kappa\tau}\,.
\end{align}

\item { $\langle\chi_{-}^{2}(t)\rangle$}:  In like manner  we find $\langle\chi_{-}^{2}(t)\rangle$,
\begin{align}
	\langle\chi_{-}^{2}(t)\rangle&=\frac{1}{m^{2}}\int_{0}^{t}\!dt'\,dt''\;d_{2}^{(-)}(t-t')d_{2}^{(-)}(t-t'')\,\langle\xi_{-}(t')\xi_{-}(t'')\rangle\notag\\
	&=\frac{2e^{2}}{m^{2}}\int_{0}^{t}\!dt'\,dt''\;d_{2}^{(-)}(t')d_{2}^{(-)}(t'')\Bigl[G_{H}(\mathbf{0},t'-t'')-G_{H}(\mathbf{z}_{1}-\mathbf{z}_{2},t'-t'')\Bigr]\,,
\end{align}
so that at late time we arrive at
\begin{align}
	\lim_{t\to\infty}\langle\chi_{-}^{2}(t)\rangle=\frac{2e^{2}}{m^{2}}\int_{-\infty}^{\infty}\!\frac{d\kappa}{2\pi}\;\lvert\overline{d}_{2}^{(-)}(\kappa)\rvert^{2}\Bigl[\overline{G}_{H}(\mathbf{0},\kappa)-\overline{G}_{H}(\mathbf{z}_{1}-\mathbf{z}_{2},\kappa)\Bigr]\,.\label{E:twiehs2}
\end{align}

\item  { $\langle\chi_{+}(t)\chi_{-}(t)\rangle$}: Finally we show that $\langle\chi_{+}(t)\chi_{-}(t)\rangle$ vanishes,
\begin{align}
	\langle\chi_{+}(t)\chi_{-}(t)\rangle&=\frac{1}{m^{2}}\int_{0}^{t}\!dt'\,dt''\;d_{2}^{(+)}(t-t')d_{2}^{(-)}(t-t'')\,\langle\xi_{+}(t')\xi_{-}(t'')\rangle=0\,,\label{E:twiehs3}
\end{align}
due to $\langle\xi_{+}(t')\xi_{-}(t'')\rangle=0$. The same holds for $\langle\chi_{-}(t)\chi_{+}(t)\rangle$, that is, $\langle\chi_{-}(t)\chi_{+}(t)\rangle=0$.
\end{itemize}

\subsection{$\dfrac{1}{2}\,\langle\{\chi_{i}(t),\chi_{j}(t)\}\rangle$ \& $\dfrac{1}{2}\,\langle\{p_{i}(t),p_{j}(t)\}\rangle$}

The displacements $\chi_{1,2}$ of both oscillators are related to $\chi_{\pm}$ by
\begin{align}
	\chi_{1}&=\chi_{+}+\frac{1}{2}\,\chi_{-}\,,&\chi_{2}&=\chi_{+}-\frac{1}{2}\,\chi_{-}\,,
\end{align} 
so with \eqref{E:twiehs1}, \eqref{E:twiehs2} and \eqref{E:twiehs3}, we can compute $\langle\chi_{1}^{2}(t)\rangle$, $\langle\chi_{2}^{2}(t)\rangle$ and $\dfrac{1}{2}\,\langle\{\chi_{1}(t),\chi_{2}(t)\}\rangle$ in the late time limit. At late times they are given by
\begin{align}
	\lim_{t\to\infty}\langle\chi_{1}^{2}(t)\rangle&=\lim_{t\to\infty}\Bigl[\langle\chi_{+}^{2}(t)\rangle+\langle\chi_{+}(t)\chi_{-}\rangle+\frac{1}{4}\,\langle\chi_{-}^{2}(t)\rangle\Bigr]=\lim_{t\to\infty}\Bigl[\langle\chi_{+}^{2}(t)\rangle+\frac{1}{4}\,\langle\chi_{-}^{2}(t)\rangle\Bigr]\,,\\
	\lim_{t\to\infty}\langle\chi_{2}^{2}(t)\rangle&=\lim_{t\to\infty}\Bigl[\langle\chi_{+}^{2}(t)\rangle-\langle\chi_{+}(t)\chi_{-}\rangle+\frac{1}{4}\,\langle\chi_{-}^{2}(t)\rangle\Bigr]=\lim_{t\to\infty}\Bigl[\langle\chi_{+}^{2}(t)\rangle+\frac{1}{4}\,\langle\chi_{-}^{2}(t)\rangle\Bigr]\,,\\
	\lim_{t\to\infty}\frac{1}{2}\,\langle\bigl\{\chi_{1}(t),\chi_{2}(t)\bigr\}\rangle&=\lim_{t\to\infty}\langle\chi_{1}(t)\chi_{2}(t)\rangle=\lim_{t\to\infty}\Bigl[\langle\chi_{+}^{2}(t)\rangle-\frac{1}{4}\,\langle\chi_{-}^{2}(t)\rangle\Bigr]\,.
\end{align}
Since $p_{\pm}=m\dot{\chi}_{\pm}$, we can easily find $\langle p_{+}^{2}(t)\rangle$, $\langle p_{-}^{2}(t)\rangle$ and $\langle p_{+}(t)p_{-}(t)\rangle$ at late time from the corresponding expressions for $\chi_{\pm}$,
\begin{align}
	\lim_{t\to\infty}\langle p_{+}^{2}(t)\rangle&=\frac{e^{2}}{2}\int_{-\infty}^{\infty}\!\frac{d\kappa}{2\pi}\;\kappa^{2}\,\lvert\overline{d}_{2}^{(+)}(\kappa)\rvert^{2}\Bigl[\overline{G}_{H}(\mathbf{0},\kappa)+\overline{G}_{H}(\mathbf{z}_{1}-\mathbf{z}_{2},\kappa)\Bigr]\,,\\
	\lim_{t\to\infty}\langle p_{-}^{2}(t)\rangle&=2e^{2}\int_{-\infty}^{\infty}\!\frac{d\kappa}{2\pi}\;\kappa^{2}\,\lvert\overline{d}_{2}^{(-)}(\kappa)\rvert^{2}\Bigl[\overline{G}_{H}(\mathbf{0},\kappa)-\overline{G}_{H}(\mathbf{z}_{1}-\mathbf{z}_{2},\kappa)\Bigr]\,,\label{E:pnfkwea}\\
	\lim_{t\to\infty}\langle p_{+}(t)p_{-}(t)\rangle&=0\,,
\end{align}
and then
\begin{align}
	\lim_{t\to\infty}\langle p_{1}^{2}(t)\rangle&=\lim_{t\to\infty}\Bigl[\langle p_{+}^{2}(t)\rangle+\frac{1}{4}\,\langle p_{-}^{2}(t)\rangle\Bigr]\,,\\
	\lim_{t\to\infty}\langle p_{2}^{2}(t)\rangle&=\lim_{t\to\infty}\Bigl[\langle p_{+}^{2}(t)\rangle+\frac{1}{4}\,\langle p_{-}^{2}(t)\rangle\Bigr]\,,\\
	\lim_{t\to\infty}\frac{1}{2}\,\langle\bigl\{ p_{1}(t),p_{2}(t)\bigr\}\rangle&=\lim_{t\to\infty}\Bigl[\langle p_{+}^{2}(t)\rangle-\frac{1}{4}\,\langle p_{-}^{2}(t)\rangle\Bigr]\,.
\end{align}
At this point it pays  for later discussions to take a closer look at $\lvert\overline{d}_{2}^{(\pm)}(\kappa)\rvert^{2}$, $\overline{G}_{H}(\mathbf{0},\kappa)$ and $\overline{G}_{H}(\mathbf{z}_{1}-\mathbf{z}_{2},\kappa)$.

We first note that $\overline{d}_{2}^{(\pm)}(\kappa)$ is related to $\widetilde{d}_{2}^{(\pm)}(s)$ in \eqref{E:qjskerw} by setting $s=-i\,\kappa$, thus  $\lvert\overline{d}_{2}^{(\pm)}(\kappa)\rvert^{2}$ are even functions of $\kappa$.  As for $\overline{G}_{H}(\mathbf{R},\kappa)$, we start with the expression of $G_{H}(\mathbf{R},\tau)$
\begin{align}
	G_{H}(\mathbf{R},\tau)&=\frac{1}{2}\int_{-\infty}^{\infty}\!\frac{d^{3}\mathbf{k}}{(2\pi)^{3}}\;\frac{1}{2\kappa}\,\coth\frac{\beta\kappa}{2}\Bigl[e^{+i\,\mathbf{k}\cdot\mathbf{R}-i\,\kappa\tau}+e^{-i\,\mathbf{k}\cdot\mathbf{R}+i\,\kappa\tau}\Bigr]\notag\\
	&=\int_{-\infty}^{\infty}\!\frac{d\kappa}{2\pi}\;\frac{1}{4\pi}\frac{\sin\kappa R}{R}\coth\frac{\beta\kappa}{2}\,e^{-i\kappa\tau}\,,
\end{align}
from which we deduced that its Fourier transform is given by
\begin{equation}
	\overline{G}_{H}(\mathbf{R},\kappa)=\frac{1}{4\pi}\frac{\sin\kappa R}{R}\coth\frac{\beta\kappa}{2}\,,
\end{equation}
with $\kappa=\lvert\mathbf{k}\rvert$ and $R=\lvert\mathbf{R}\rvert$, and that $\overline{G}_{H}(\mathbf{R},\kappa)$ is also an even function of $\kappa$. In fact, by the fluctuation-dissipation theorem
\begin{equation}
	\overline{G}_{H}(\mathbf{R},\kappa)=\frac{1}{4\pi}\frac{\sin\kappa R}{R}\coth\frac{\beta\kappa}{2}=\coth\frac{\beta\kappa}{2}\,\operatorname{Im}\overline{G}_{R}(\mathbf{R},\kappa)\,,
\end{equation}
all uncertainty functions of the canonical variables for the CoM/relative mode can be further simplified. Generically, for example, the uncertainty of $\chi$ at late time reduces to 
\begin{align}
	\langle\chi^{2}(\infty)\rangle=c\,\frac{e^{2}}{m^{2}}\int_{-\infty}^{\infty}\!\frac{d\kappa}{2\pi}\;\lvert\overline{d}_{2}(\kappa)\rvert^{2}\,\overline{G}_{H}(\kappa)&=c\,\frac{e^{2}}{m^{2}}\int_{-\infty}^{\infty}\!\frac{d\kappa}{2\pi}\;\frac{\coth\frac{\beta\kappa}{2}\,\operatorname{Im}\overline{G}_{R}(\kappa)}{\lvert-\kappa^{2}-\frac{e^{2}}{m}\,\overline{G}_{R}(\kappa)+\omega^{2}\rvert^{2}}\notag\\
	&=\frac{2c}{m}\operatorname{Im}\int_{0}^{\infty}\!\frac{d\kappa}{2\pi}\;\overline{d}_{2}(\kappa)\,\coth\frac{\beta\kappa}{2}\,,\label{E:dersdsjk}
\end{align}
where $c=1/2$ for the CoM mode and $c=2$ for the relative mode, and we have used the fact that 
\begin{equation}
	\overline{d}_{2}(\kappa)=\frac{1}{-\kappa^{2}-\frac{e^{2}}{m}\,\overline{G}_{R}(\kappa)+\omega^{2}}\,.
\end{equation}
Here $\overline{G}$ contains both $\overline{G}(\mathbf{0},\kappa)$ and $\overline{G}(\mathbf{R},\kappa)$. Likewise, the late-time uncertainty of $p$ can be written as
\begin{equation}\label{E:dersdsjj}
	\langle p^{2}(\infty)\rangle=2mc\,\operatorname{Im}\int_{0}^{\infty}\!\frac{d\kappa}{2\pi}\;\kappa^{2}\,\overline{d}_{2}(\kappa)\,\coth\frac{\beta\kappa}{2}\,.
\end{equation} 
Eqs.~\eqref{E:dersdsjk} and \eqref{E:dersdsjj} apply to both the CoM and the relative modes of the oscillators.

\subsection{$\frac{1}{2}\,\langle\{\chi_{i}(t),p_{j}(t)\}\rangle$}

As for the covariance matrix elements like $\frac{1}{2}\,\langle\{\chi_{i}(t),p_{j}(t)\}\rangle$, we have
\begin{align}
	\frac{1}{2}\,\langle\bigl\{\chi_{1}(t),p_{1}(t)\bigr\}\rangle&=\langle\chi_{+}(t)p_{+}(t)\rangle+\frac{1}{2}\,\langle\chi_{-}(t)p_{+}(t)\rangle+\frac{1}{2}\,\langle\chi_{+}(t)p_{-}(t)\rangle+\frac{1}{4}\,\langle\chi_{-}(t)p_{-}(t)\rangle\,,\\
	\frac{1}{2}\,\langle\bigl\{\chi_{2}(t),p_{2}(t)\bigr\}\rangle&=\langle\chi_{+}(t)p_{+}(t)\rangle-\frac{1}{2}\,\langle\chi_{-}(t)p_{+}(t)\rangle-\frac{1}{2}\,\langle\chi_{+}(t)p_{-}(t)\rangle+\frac{1}{4}\,\langle\chi_{-}(t)p_{-}(t)\rangle\,,\\
	\frac{1}{2}\,\langle\bigl\{\chi_{1}(t),p_{2}(t)\bigr\}\rangle&=\langle\chi_{+}(t)p_{+}(t)\rangle+\frac{1}{2}\,\langle\chi_{-}(t)p_{+}(t)\rangle-\frac{1}{2}\,\langle\chi_{+}(t)p_{-}(t)\rangle-\frac{1}{4}\,\langle\chi_{-}(t)p_{-}(t)\rangle\,,\\
	\frac{1}{2}\,\langle\bigl\{\chi_{2}(t),p_{1}(t)\bigr\}\rangle&=\langle\chi_{+}(t)p_{+}(t)\rangle-\frac{1}{2}\,\langle\chi_{-}(t)p_{+}(t)\rangle+\frac{1}{2}\,\langle\chi_{+}(t)p_{-}(t)\rangle-\frac{1}{4}\,\langle\chi_{-}(t)p_{-}(t)\rangle\,.
\end{align}
At late time $t\to\infty$ we readily find
\begin{align}
	\langle\chi_{+}(t)p_{+}(t)\rangle&=\frac{1}{m}\int_{0}^{t}\!dt'dt''\;d_{2}^{(+)}(t-t')\dot{d}_{2}^{(+)}(t-t'')\,\langle\xi_{+}(t')\xi_{+}(t'')\rangle\notag\\
	&=\frac{e^{2}}{2m}\int_{0}^{t}\!dt'dt''\;d_{2}^{(+)}(t')\dot{d}_{2}^{(+)}(t'')\Bigl[G_{H}(\mathbf{0},t'-t'')+G_{H}(\mathbf{z}_{1}-\mathbf{z}_{2},t'-t'')\Bigr]\notag\\
t\to\infty\quad&\to i\,\frac{e^{2}}{2m}\int_{-\infty}^{\infty}\!\frac{d\kappa}{2\pi}\;\kappa\,\lvert\overline{d}_{2}^{(+)}(\kappa)\rvert^{2}\Bigl[\overline{G}_{H}(\mathbf{0},\kappa)+\overline{G}_{H}(\mathbf{z}_{1}-\mathbf{z}_{2},\kappa)\Bigr]\,.
\end{align}
Since $\lvert\overline{d}_{2}^{(+)}(\kappa)\rvert^{2}$, $\overline{G}_{H}(\mathbf{0},\kappa)$ and $\overline{G}_{H}(\mathbf{z}_{1}-\mathbf{z}_{2},\kappa)$ are all even functions of $\kappa$, we conclude that
\begin{equation}
	\langle\chi_{+}(t)p_{+}(t)\rangle=0\,.
\end{equation} 
Next we examine $\langle\chi_{-}(t)p_{-}(t)\rangle$ and find that
\begin{align}
	\langle\chi_{-}(t)p_{-}(t)\rangle&=\frac{1}{m}\int_{0}^{t}\!dt'dt''\;d_{2}^{(-)}(t-t')\dot{d}_{2}^{(-)}(t-t'')\,\langle\xi_{-}(t')\xi_{-}(t'')\rangle\notag\\
	&=\frac{2e^{2}}{m}\int_{0}^{t}\!dt'dt''\;d_{2}^{(-)}(t')\dot{d}_{2}^{(-)}(t'')\Bigl[G_{H}(\mathbf{0},t'-t'')-G_{H}(\mathbf{z}_{1}-\mathbf{z}_{2},t'-t'')\Bigr]\notag\\
t\to\infty\quad&\to i\,\frac{2e^{2}}{m}\int_{-\infty}^{\infty}\!\frac{d\kappa}{2\pi}\;\kappa\,\lvert\overline{d}_{2}^{(-)}(\kappa)\rvert^{2}\Bigl[\overline{G}_{H}(\mathbf{0},\kappa)-\overline{G}_{H}(\mathbf{z}_{1}-\mathbf{z}_{2},\kappa)\Bigr]\notag\\
	&=0\,.
\end{align}
Finally we calculate $\langle\chi_{+}(t)p_{-}(t)\rangle$ and $\langle\chi_{-}(t)p_{+}(t)\rangle$,
\begin{align}
	\langle\chi_{+}(t)p_{-}(t)\rangle&=\frac{1}{m}\int_{0}^{t}\!dt'dt''\;d_{2}^{(+)}(t-t')\dot{d}_{2}^{(-)}(t-t'')\,\langle\xi_{+}(t')\xi_{-}(t'')\rangle\notag\\
	&=0\,,
\end{align}
because $\langle\xi_{+}(t')\xi_{-}(t'')\rangle=0$. Thus all of the cross-correlations between $\chi_{i}$ and $p_{j}$ vanish at late times, and the only remaining nonvanishing elements of the covariance matrix in the limit $t\to\infty$ are $\langle\chi_{1,2}^{2}(t)\rangle$, $\dfrac{1}{2}\,\langle\{\chi_{1}(t),\chi_{2}(t)\}\rangle$, $\langle p_{1,2}^{2}(t)\rangle$ and $\dfrac{1}{2}\,\langle\{p_{1}(t),p_{2}(t)\}\rangle$. These nonvanishing elements of the covariance matrix can be assembled from the expectation values of $\langle\chi_{\pm}^{2}(t)\rangle$ and $\langle p_{\pm}^{2}(t)\rangle$. Their computations are shown in the Appendix~\ref{E:nvndfkse}.\\


\section{Parameter Constraints}

To relate to realistic physical conditions it is important to ascertain the range of validity of the most relevant parameters in this setup,  such as the separation $\ell$ between the oscillators, the damping constant $\gamma$, and the inter-oscillator coupling $\sigma$.

We have two oscillators residing in a common bath, which is described by a quantum scalar field initially in its thermal state, so they will experience different but correlated field strengths. If we move them closer to one another, we find in \eqref{E:qarens} that when $\omega_{\pm}\ell\ll1$, their equations of motion correspond to a configuration in which two oscillators seem to be placed at the same spatial location. This is a common configuration used for the investigation of entanglement dynamics \cite{Goan,Paz,HB08}.  However if we move the oscillators closer to one another, instability in their motion can occur. \\

\paragraph{Close Proximity Instablity} 

As discussed in the paragraph below \eqref{E:woijeow}, the separation $\ell$ must be greater than $2\gamma/\omega_{\pm}^{2}$; otherwise the equations of motion will have a runaway solution. As seen from the criterion, this instability can also occur for the strong damping case with a moderate separation between oscillators, i.e. $\dfrac{\gamma}{\omega_{\pm}}>\omega_{\pm}\ell$. In fact if we write the stability criterion explicitly in terms of the coupling constants, 
\begin{equation}
	\ell>\frac{2\gamma}{\omega^{2}\pm\sigma}\,,
\end{equation}
we may easily conclude that the minimum separation to avoid the instability increases with stronger damping and stronger inter-oscillator coupling. In other words, strong coupling,  either between the oscillator and the environment or between the oscillators, tends to induce unstable motion of the oscillator in the current configuration.  It is also obvious that the slow mode places a stricter constraint on the minimum separation to ensure a stable evolution of the oscillators' motion. This is a distinct feature of the non-Markovian dynamics of the detectors in a shared bath configuration,   in that the strong damping in the Markovian dynamics in the private-bath configuration only results in overdamped motion.

If one is interested in how thermal entanglement can be sustained for a finite separation between the two detectors in a common bath,  the $\omega_{\pm}\ell\ll1$ case is not of special interest because  the two detectors essentially merge together. The more relevant case would be for $\omega_{\pm}\ell\not\ll1$.  \\

\paragraph{Dynamics at late times}

Based on the previous considerations, the perturbative solution to $\widetilde{g}_{\pm}(s)=0$ will give the dominant contribution at late time and the approximation will be improved with smaller ratio of $\gamma/(\omega_{\pm}^{2}\ell)$. In this approximation, all the non-Markovian effects between the oscillators are summarized into the effective damping $\Gamma_{\pm}$ and the effective natural frequency $W_{\pm}$,
\begin{align*}
	\Gamma_{\pm}&=\gamma\left[1\pm\frac{\sin\omega_{\pm}\ell}{\omega_{\pm}\ell}\right]\,,&W_{\pm}&=\omega_{\pm}\left[1\mp\frac{\gamma}{\omega_{\pm}}\frac{\cos\omega_{\pm}\ell}{\omega_{\pm}\ell}\right]\,.
\end{align*}
of the oscillators,  whence the oscillators superficially undergo Markovian dynamics. All the information about the history or the finite separation is hidden in these effective parameters. This may not be too surprising because the small ratio $\gamma/(\omega_{\pm}^{2}\ell)$ implies short memory lapse.  At each exchange of the mutual influence between the oscillators, the non-Markovian effect will be suppressed by that factor; thus it will not last more than a few exchanges. This also implies that the initial information hardly survives at late time. That is why the dynamics of the oscillators in these limits looks so Markovian.


\section{Entanglement Measure}

We will use the negativity as the entanglement measure because it offers unambiguous quantification of entanglement for a symmetric two-mode Gaussian state. It is defined in terms of the symplectic eigenvalue $\eta_{<}$ of the partially-transposed covariance matrix
\begin{align}\label{E:dfkefjdkw}
	\mathcal{N}(\rho)&=\max\bigl\{0,\frac{1-2\eta_{<}}{2\eta_{<}}\bigr\}\,,&E_{\mathcal{N}}(\rho)&=\max\bigl\{0,-\ln 2\eta_{<}\bigr\}\,,
\end{align}
where $\eta_{<}$ is the smaller of the pair of symplectic eigenvalues ($\eta_{>}$, $\eta_{<}$) of the partially-transposed covariance matrix $\mathbf{V}^{pt}$. \textit{Entanglement occurs when $\eta_{<}<1/2$, and the degree of entanglement is described by the negativity}.

\subsection{Negativity in terms of symplectic eigenvalue $\eta_<$}

The covariance matrix $\mathbf{V}$ of a two-mode state is define by
\begin{equation}
	\mathbf{V}=\frac{1}{2}\operatorname{Tr}\Bigl[\rho\bigl\{\mathbf{X},\mathbf{X}^{T}\bigr\}\Bigr]\,,
\end{equation}
where $\mathbf{X}^{T}=(\chi_{1},p_{1},\chi_{2},p_{2})$ and $\rho$ is the density matrix of the two-mode state. Thus the covariance matrix of a two-mode state is a $4\times4$ matrix, which consists of the uncertainties or the cross-correlations between the canonical variables of the state. The separability criterion or the entanglement measure based on the covariance matrix is particularly convenient and calculable for Gaussian continuous variable systems because its covariance matrix has a finite dimension, in comparison with the infinite-dimensional density matrix, which is commonly used in discrete systems. The partial transpose of a covariance matrix can be simply constructed  by changing the sign of one of the momentum variable in the original covariance matrix. The symplectic eigenvalues of the partially-transposed covariance matrix $\mathbf{V}^{pt}$ can thus be found by taking the absolute value of the ordinary eigenvalues of the matrix $i\,\mathbf{J}^{\oplus^{2}}\cdot\mathbf{V}^{pt}$, where 
\begin{equation}
	\mathbf{J}=\begin{pmatrix}0&+1\\-1&0\end{pmatrix}\,.
\end{equation}
If we write the covariance matrix in a block form
\begin{equation}
	\mathbf{V}=\begin{pmatrix}\mathbf{A} &\mathbf{C} \\
					\mathbf{C}^{T} &\mathbf{B}
				\end{pmatrix}\,,
\end{equation}
then the symplectic eigenvalues $\eta_{\gtrless}$ of the partially-transposed covariance matrix $\mathbf{V}^{pt}$ can be expressed in terms of these $2\times2$ matrices $\mathbf{A}$, $\mathbf{B}$, $\mathbf{C}$ as 
\begin{align}\label{E:bheuhs}
	\eta_{\gtrless}&=\left[\left(\frac{\det\mathbf{A}+\det\mathbf{B}}{2}-\det\mathbf{C}\right)\pm\sqrt{\left(\frac{\det\mathbf{A}+\det\mathbf{B}}{2}-\det\mathbf{C}\right)^{2}-\det\mathbf{V}}\right]^{\frac{1}{2}}\,,
\end{align}
where alternatively $\det\mathbf{V}$ can be written as $\det\mathbf{A}\,\det\mathbf{B}+(\det\mathbf{C})^{2}-\operatorname{Tr}\{\mathbf{A}\cdot\mathbf{J}\cdot\mathbf{C}\cdot\mathbf{J}\cdot\mathbf{B}\cdot\mathbf{J}\cdot\mathbf{C}^{T}\cdot\mathbf{J}\}$.

In the case that $\mathbf{A}=\mathbf{B}$ and the matrix $\mathbf{C}$ is diagonal, the symplectic eigenvalue $\eta_{<}$ of $\mathbf{V}^{pt}$ takes a particularly neat form
\begin{equation}\label{E:dbekrheq}
	\eta_{<}^{2}=\mathcal{V}_{11}\mathcal{V}_{22}-\mathcal{V}_{13}\mathcal{V}_{24}-\lvert\mathcal{V}_{22}\mathcal{V}_{13}-\mathcal{V}_{11}\mathcal{V}_{24}\rvert=\begin{cases}
								\bigl(\mathcal{V}_{11}-\mathcal{V}_{13}\bigr)\bigl(\mathcal{V}_{22}+\mathcal{V}_{24}\bigr)\,,&\mathcal{V}_{22}\mathcal{V}_{13}>\mathcal{V}_{11}\mathcal{V}_{24}\,,\\
								\bigl(\mathcal{V}_{11}+\mathcal{V}_{13}\bigr)\bigl(\mathcal{V}_{22}-\mathcal{V}_{24}\bigr)\,,&\mathcal{V}_{22}\mathcal{V}_{13}<\mathcal{V}_{11}\mathcal{V}_{24}\,,
							\end{cases}
\end{equation}
with
\begin{align*}
	\mathcal{V}_{11}&=\langle\chi_{1}^{2}(t)\rangle\,,&\mathcal{V}_{22}&=\langle p_{1}^{2}(t)\rangle\,,&\mathcal{V}_{13}&=\frac{1}{2}\,\langle\bigl\{\chi_{1}(t),\chi_{2}(t)\bigr\}\rangle\,,&\mathcal{V}_{24}&=\frac{1}{2}\,\langle\bigl\{p_{1}(t),p_{2}(t)\bigr\}\rangle\,.
\end{align*}
We readily see that
\begin{align}
	\mathcal{V}_{11}\mp\mathcal{V}_{13}&=\frac{1}{2}\,\langle\bigl\{\chi_{1},\chi_{1}\mp\chi_{2}\bigr\}\rangle=\begin{cases}\dfrac{1}{2}\,\langle\chi_{-}^{2}\rangle\,,&-\,,\vspace{9pt}\\2\,\langle\chi_{+}^{2}\rangle\,,&+\,,\end{cases}\\
	\mathcal{V}_{22}\pm\mathcal{V}_{24}&=\frac{1}{2}\,\langle\bigl\{p_{1},p_{1}\pm p_{2}\bigr\}\rangle=\begin{cases}2\,\langle p_{+}^{2}\rangle\,,&+\,,\vspace{9pt}\\\dfrac{1}{2}\,\langle p_{-}^{2}\rangle\,,&-\,,\end{cases}
\end{align}
are associated with the dynamics of the normal modes of the joint system. Thus depending on the sign of $\mathcal{V}_{22}\mathcal{V}_{13}-\mathcal{V}_{11}\mathcal{V}_{24}$ the symplectic eigenvalue $\eta_{<}$ will take different forms
\begin{equation}\label{E:qhbjdhs}
	\eta_{<}^{2}=\begin{cases}
					\langle\chi_{-}^{2}\rangle\langle p_{+}^{2}\rangle\,,&\mathcal{V}_{22}\mathcal{V}_{13}>\mathcal{V}_{11}\mathcal{V}_{24}\,,\\
					\langle\chi_{+}^{2}\rangle\langle p_{-}^{2}\rangle\,,&\mathcal{V}_{22}\mathcal{V}_{13}<\mathcal{V}_{11}\mathcal{V}_{24}\,.
				\end{cases}
\end{equation}
Likewise we find
\begin{equation}\label{E:qhbjdht}
	\eta_{>}^{2}=\begin{cases}
					\langle\chi_{+}^{2}\rangle\langle p_{-}^{2}\rangle\,,&\mathcal{V}_{22}\mathcal{V}_{13}>\mathcal{V}_{11}\mathcal{V}_{24}\,,\\
					\langle\chi_{-}^{2}\rangle\langle p_{+}^{2}\rangle\,,&\mathcal{V}_{22}\mathcal{V}_{13}<\mathcal{V}_{11}\mathcal{V}_{24}\,.
				\end{cases}
\end{equation}
In fact we observe that 
\begin{align}
	\mathcal{V}_{22}\mathcal{V}_{13}-\mathcal{V}_{11}\mathcal{V}_{24}&=\frac{1}{2}\Bigl[\langle\chi_{+}^{2}\rangle\langle p_{-}^{2}\rangle-\langle\chi_{-}^{2}\rangle\langle p_{+}^{2}\rangle\Bigr]\,,
\end{align}
so \eqref{E:qhbjdhs} can be summarized into
\begin{align}\label{E:qejdla}
	\eta_{<}^{2}&=\min\bigl\{\langle\chi_{+}^{2}\rangle\langle p_{-}^{2}\rangle,\,\langle\chi_{-}^{2}\rangle\langle p_{+}^{2}\rangle\bigr\}\,,&\eta_{>}^{2}&=\max\bigl\{\langle\chi_{+}^{2}\rangle\langle p_{-}^{2}\rangle,\,\langle\chi_{-}^{2}\rangle\langle p_{+}^{2}\rangle\bigr\}\,.
\end{align}
This distinction can be handy in interpreting entanglement. Furthermore,  in either case, for the entanglement to exist, we would like to have the uncertainties of the corresponding canonical variables to be as small as possible such that $\eta_{<}^{2}$ can be smaller than $1/4$.
\begin{figure}
	\centering
    \scalebox{0.6}{\includegraphics{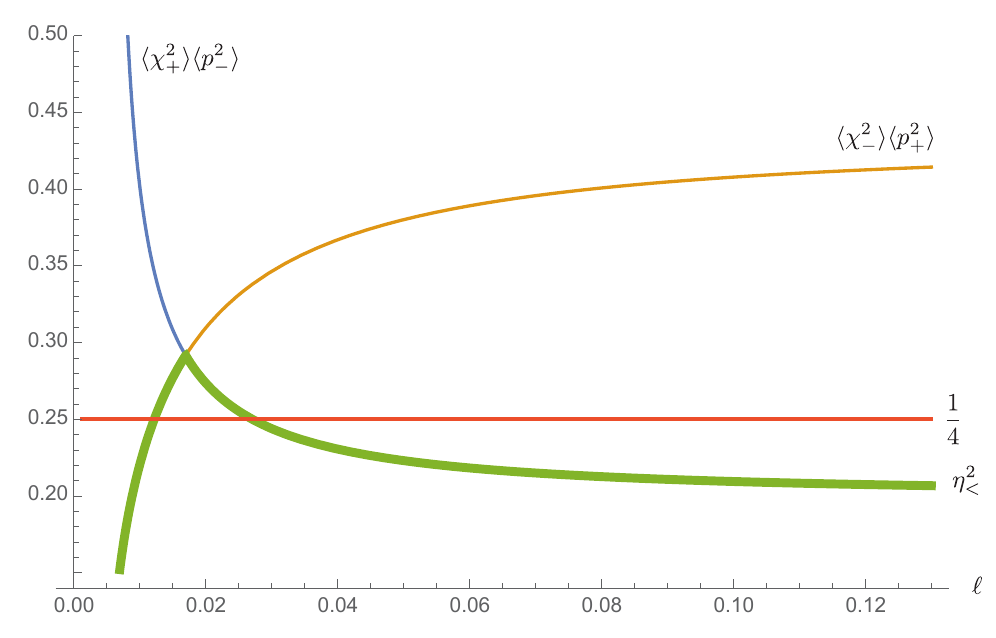}}
    \captionsetup{justification=raggedright}
    \caption{Behavior of $\eta_{<}^{2}$ as a function of separation $\ell$ between the two oscillators. The symplectic eigenvalue $\eta_{<}$ of the partially transposed covariance matrix $\mathbf{V}^{pt}$ can be used to quantify the entanglement of a symmetric Gaussian state, described by $\mathbf{V}$. The value of $\eta_{<}^{2}$ is given by the minimum between $\langle\chi_{+}^{2}\rangle\langle p_{-}^{2}\rangle$, and $\langle\chi_{-}^{2}\rangle\langle p_{+}^{2}\rangle$. When $\eta_{<}^{2}$ is less than $1/4$, the state is entangled; otherwise it is separable. There may be two different critical separation where the entanglement disappears. The point where the curves of $\langle\chi_{+}^{2}\rangle\langle p_{-}^{2}\rangle$, and $\langle\chi_{-}^{2}\rangle\langle p_{+}^{2}\rangle$ intersect always corresponds to a separable state.}\label{Fi:etasml}
\end{figure}

\subsection{Interpretation of Entanglement Measure in Terms of Effective Frequencies}

Since the symplectic eigenvalue $\eta_{<}$ is constructed out of $\langle\chi_{\pm}^{2}\rangle$ and $\langle p_{\pm}^{2}\rangle$, let us work out their general behaviors with respect to the parameters at hand. Recall the equations of motion for the faster/slow modes are given by
\begin{equation}
	\ddot{\chi_{\pm}}+2\gamma\,\dot{\chi}_{\pm}(s)\mp2\gamma\,\frac{\theta(s-\ell)}{\ell}\,\chi_{\pm}(s-\ell)+\omega^{2}_{\pm}\,\chi_{\pm}(s)=\frac{1}{m}\,\xi_{\pm}(s)\,.
\end{equation}
From their frequency representations, 
\begin{align}\label{E:xmsdhs}
	\bigl[-\kappa^{2}-i\,2\gamma\kappa+\bigl(\omega^{2}\pm\sigma\bigr)\mp\frac{2\gamma}{\ell}\,e^{i\,\kappa\ell}\bigr]\overline{\chi}_{\pm}(\kappa)&=\frac{1}{m}\,\overline{\xi}_{\pm}(\kappa)\,,&&\Leftrightarrow&\overline{\chi}_{\pm}(\kappa)&=\frac{1}{m}\,\overline{d}_{2}^{(\pm)}(\kappa)\overline{\xi}_{\pm}(\kappa)\,,
\end{align}
we may attempt to break the $e^{i\kappa\ell}$ into its real and imaginary parts, and group them with the oscillating frequency and the damping constant respectively to form the \textit{effective oscillating frequency}
\begin{equation}\label{E:xegsds}
	W_{\pm}^{2}=\omega^{2}\pm\sigma\mp\frac{2\gamma}{\ell}\,\cos\kappa\ell\,,
\end{equation}
and the \textit{effective damping constants}
\begin{equation}
	\Gamma_{\pm}=\gamma\pm\frac{\gamma}{\kappa\ell}\,\sin\kappa\ell\,.
\end{equation}
In so doing heuristically speaking we have two effective uncoupled, damped, driven oscillators. For weak oscillator-bath coupling $\gamma$, the curve of $\overline{d}_{2}^{(\pm)}(\kappa)$ has a very narrow resonance peak at about $\kappa=\omega_{\pm}$. Thus we may replace $\kappa$ in $\Gamma_{\pm}$ and $W_{\pm}^{2}$ with its typical value, say $\omega_{\pm}$. However, from hindsight we find this choice is not adequate for $W_{\pm}$. Compared with the result of $\langle\chi_{\pm}^{2}(\infty)\rangle$ at zero temperature, the better option of the effective oscillating frequency will be
\begin{equation}\label{E:xegsd} 
	W_{\pm}^{2}=\omega^{2}\pm\sigma\mp\frac{2\gamma}{\ell}\,,
\end{equation}
in order to avoid wiggling artifact due to the $\cos$ expression. Now based on the viewpoint of the effective oscillators, we expect that when the temperature is not too high, the uncertainty of the corresponding canonical variables should be about the order (apart from the mass scale)
\begin{align}
	\langle\chi_{\pm}^{2}\rangle&\sim\mathcal{O}(\frac{1}{W_{\pm}})\,,&\langle p_{\pm}^{2}\rangle&\sim\mathcal{O}(W_{\pm})\,,
\end{align}
and on the other hand at high temperature their values are dominated by the temperature, leading to
\begin{align}\label{E:zordfa}
	\langle\chi_{\pm}^{2}\rangle&\sim\mathcal{O}(\frac{1}{\beta W^{2}_{\pm}})\,,&\langle p_{\pm}^{2}\rangle&\sim\mathcal{O}(\frac{1}{\beta})\,.
\end{align}
Let us for the moment forget about any intricacy due to the cutoff scale because it does not really depend on any other parameters in the setup.\\

\paragraph*{Case  $W_{+}>W_{-}$}

With these understandings, when $W_{+}>W_{-}$ and the temperature of the bath is not high, we have
\begin{align}
	\langle\chi_{+}^{2}\rangle&<\langle\chi_{-}^{2}\rangle\,,&&\text{and}&\langle p_{-}^{2}\rangle&<\langle p_{+}^{2}\rangle\,,&&\Rightarrow&\langle\chi_{+}^{2}\rangle\langle p_{-}^{2}\rangle&<\langle\chi_{-}^{2}\rangle\langle p_{+}^{2}\rangle\,.
\end{align}
From \eqref{E:qejdla} the smaller symplectic eigenvalue $\eta_{<}$ of the covariance matrix in this case will be given by
\begin{align}
	\eta_{<}^{2}&=\langle\chi_{+}^{2}\rangle\langle p_{-}^{2}\rangle\sim\mathcal{O}(\frac{W_{-}}{W_{+}})\,,&&\text{when $W_{+}>W_{-}$}\,.
\end{align}
With the help of the definition of the effective frequencies \eqref{E:xegsd}, we see that for a fixed $\omega$ the effective frequencies tend to take the opposite trends, that is, when $W_{+}$ gets bigger, $W_{-}$ becomes smaller, and vice versa. It then implies that $\eta_{<}^{2}$ tends to decrease with larger values of $W_{+}$; thus it is more likely  that entanglement can exist for larger $W_{+}$. More precisely speaking, as seen from \eqref{E:xegsd} with a fixed $\omega$, the effective oscillating frequency of the CoM mode $W_{+}$ tends to increase with either larger inter-oscillator coupling $\sigma$ or larger separation $\ell$ between oscillators. However in this case, the effect of the oscillator separation is usually relatively small, overshadowed by the effect due to the inter-oscillator coupling, because $W_{+}>W_{-}$ already implies that $\sigma>2\gamma/\ell$. \textit{The inter-oscillator coupling will play the dominant role in sustaining entanglement,} if it exists.  Therefore stronger inter-oscillator coupling will benefit entanglement when both oscillators are far apart. \\

\paragraph*{Case $W_{-}>W_{+}$}

For the opposite case $W_{-}>W_{+}$, in the low temperature regime, we have 
\begin{align}
	\langle\chi_{-}^{2}\rangle&<\langle\chi_{+}^{2}\rangle\,,&&\text{and}&\langle p_{+}^{2}\rangle&<\langle p_{-}^{2}\rangle\,,&&\Rightarrow&\langle\chi_{-}^{2}\rangle\langle p_{+}^{2}\rangle&<\langle\chi_{+}^{2}\rangle\langle p_{-}^{2}\rangle\,.
\end{align}
The symplectic eigenvalue $\eta_{<}$ is then given by
\begin{align}
	\eta_{<}^{2}&=\langle\chi_{-}^{2}\rangle\langle p_{+}^{2}\rangle\,,&&\text{when $W_{-}>W_{+}$}\,.
\end{align}
Following the same arguments,  larger values of $W_{-}$ will allow entanglement to better survive at late time. This implies lowering the inter-oscillator coupling strength or shortening the separation between the oscillators can raise the value of $W_{-}$, thus increase the chance that  entanglement can exist. In this case $W_{-}>W_{+}$ leads to $\sigma<2\gamma/\ell$, so the effect of the inter-oscillator coupling becomes subdominant.  {The separation between the oscillators plays a more important role}. \textit{The entanglement is more likely to survive for shorter oscillator separation in the case of vanishing inter-oscillator coupling}. \\

\paragraph*{Sign Switch in $W_{+}-W_{-}$}

Following the previous arguments and the definition of the effective frequencies, we see when $\sigma$ is roughly equal to $2\gamma/\ell$, the difference $W_{+}-W_{-}$ can switch sign. There the curves of $\langle\chi_{+}^{2}\rangle\langle p_{-}^{2}\rangle$ and $\langle\chi_{-}^{2}\rangle\langle p_{+}^{2}\rangle$ will cross over, so the symplectic eigenvalue $\eta_{<}$ will change from one form to the other. In addition, the symplectic eigenvalues of the covariance matrix can be degenerate in this regime. Since the separability criterion $\mathbf{V}^{pt}+i\,\pmb{\Omega}/2\geq0$ requires
\begin{align}\label{E:xxnekrhe}
	\bigl(\eta_{>}^{2}-\frac{1}{4}\bigr)\bigl(\eta_{<}^{2}-\frac{1}{4}\bigr)&\geq0\,,
\end{align}
the degeneracy of the symplectic eigenvalues implies
\begin{align}\label{E:yynekrhe}
	\bigl(\eta_{>}^{2}-\frac{1}{4}\bigr)\bigl(\eta_{<}^{2}-\frac{1}{4}\bigr)=\bigl(\eta^{2}-\frac{1}{4}\bigr)^{2}
\end{align}
with the degenerate value $\eta_{>}=\eta_{<}=\eta$. Eq.~\eqref{E:yynekrhe} is always greater than or equal to zero, so when $\sigma\sim2\gamma/\ell$ the system always remain in a separable state.\\

\paragraph*{High Temperature Regime}

Finally in the high temperature regime, in general $\langle\chi_{\pm}^{2}\rangle$ and $\langle p_{\pm}^{2}\rangle$ are very large because they are all proportional to the temperature, as seen from \eqref{E:zordfa}. Hence it is very difficult to maintain either $\langle\chi_{+}^{2}\rangle\langle p_{-}^{2}\rangle$ or $\langle\chi_{-}^{2}\rangle\langle p_{+}^{2}\rangle$ smaller than $1/4$. This explains why in the configuration studied here, namely two detectors in a common bath,  entanglement disappears at high temperature.  Eq.~\eqref{E:zordfa} also tells that since the the product $\langle\chi_{+}^{2}\rangle\langle p_{-}^{2}\rangle$ or $\langle\chi_{-}^{2}\rangle\langle p_{+}^{2}\rangle$ is about the order
\begin{equation}
	\eta_{<}^{2}\sim\langle\chi_{\pm}^{2}\rangle\langle p_{\mp}^{2}\rangle\sim\mathcal{O}(\frac{1}{\beta^{2}W_{\pm}^{2}})\,,
\end{equation} 
\textit{the critical temperature of the entanglement will be at most of the order
}\begin{equation}
	\beta_{c}W_{\pm}\sim\mathcal{O}(1)\,.
\end{equation}
Higher temperature will tend to render the quantity $\beta W_{\pm}$ too small to make $\eta_{<}^{2}$ {greater} than $1/4$. Thus entanglement may not exist any more at higher temperature. Moreover, it implies that when $W_{+}>W_{-}$ the critical temperature $\beta_{c}^{-1}$ is at most about the order $W_{+}$, while in the case $W_{-}>W_{+}$, the critical temperature is of the order $W_{-}$.

These are the general  characteristics of quantum entanglement in a common bath.  We now proceed to a more detailed quantitative analysis.

\section{Late-time Entanglement Analysis}
The above qualitative description shows the richness  of the entanglement  behavior at late times (much greater than the relaxation time $\gamma^{-1}$) arising from the interplay between various physical scales. Here we shall examine the zero temperature case and identify more precisely the aforementioned features. The results for the low and high temperature cases will be reported  in a separate paper \cite{HHQEnt1}.  

Owing to the multitude of length scales in question,  we first address the choices of the ranges of length scales involved in Sec.~\ref{S:gdngdkfgn1}. We then analyze the critical separation for the regime  the direct coupling dominates in Sec.~\ref{S:gdngdkfgn2} and for the regime the non-Markovian effects govern in Sec.~\ref{S:gdngdkfgn4}.

\subsection{Identification of Relevant Scales}\label{S:gdngdkfgn1}
We suppose that within the relaxation time of the system, mutual influence between the  two oscillators, in terms of retardation radiation, has occurred sufficiently many times.  It does not take long for weak bath-oscillator couplings.  This implies $\gamma^{-1}>\ell$. Depending on $\sigma\gtrless\omega^{2}/2$, we have
\begin{align}
	\sigma&>\frac{\omega^{2}}{2}\,,&&\Rightarrow&\omega_{-}^{2}&<\frac{\omega^{2}}{2}<\sigma\,,&&\Rightarrow&\frac{2\gamma}{\omega_{-}^{2}\ell}&>\frac{4\gamma}{\omega^{2}\ell}>\frac{2\gamma}{\sigma\ell}\,,\\
	\sigma&<\frac{\omega^{2}}{2}\,,&&\Rightarrow&\omega_{-}^{2}&>\frac{\omega^{2}}{2}>\sigma\,,&&\Rightarrow&\frac{2\gamma}{\omega_{-}^{2}\ell}&<\frac{4\gamma}{\omega^{2}\ell}<\frac{2\gamma}{\sigma\ell}\,.
\end{align}
Thus when $\varsigma>1$, the stability criterion is automatically satisfied if $\sigma<\omega^{2}/2$, but it has to be enforced if $\sigma>\omega^{2}/2$. On the other hand, when $\varsigma<1$, the stability criterion can be violated when $\sigma>\omega^{2}/2$, but it still needs to be enforced if $\sigma<\omega^{2}/2$. Following our previous discussions, we know that $\varsigma\simeq\mathcal{O}(1)$ is the region where the ``phase transition'' of $\eta_{<}^{2}$ occurs and the dominance between the direct and indirect interactions between the oscillators swaps. When the inter-oscillator coupling $\sigma$ is greater than $\omega^{2}/2$,  our previous result implies that it is hardly possible to sustain late time entanglement for separations shorter than $2\gamma/\sigma$ because the instability may spoil everything.

Even though we require $\dfrac{2\gamma}{\omega_{-}^{2}\ell}<1$ for the sake of stability and convergence of perturbative calculation, the choice of $\omega_{-}\ell$ can  be ambiguous because there is no restriction on it. For \textit{weak oscillator-bath coupling} $\gamma/\omega_{-}<1$ we still have the freedom to choose $\omega_{-}\ell\gtrless1$. We will return to this discussion later.  In addition because $\gamma/\omega_{-}<1$, we may not bother to distinguish the resonance frequencies of the normal modes $\Omega_{\pm}$ from $\omega_{\pm}$ unless confusion exists.

\begin{figure}
	\centering
	\scalebox{0.5}{\includegraphics{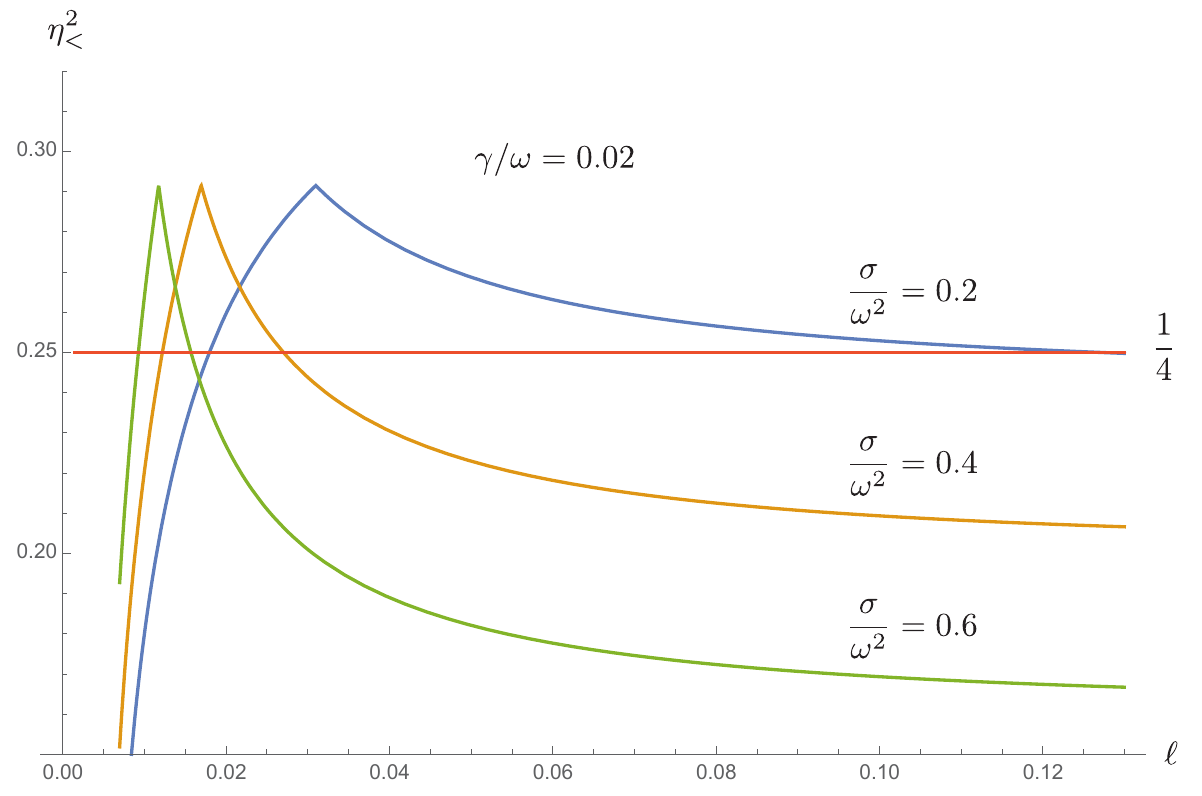}}
    \captionsetup{justification=raggedright}
    \caption{For sufficiently strong inter-oscillator coupling strength $\sigma$, the oscillators may remain entangled even when they are far apart. This late time entanglement deteriorates with shorter separation when $\ell>\mathcal{O}(2\gamma/\sigma)$. Disentanglement will occur when the distance between oscillators is less than the critical separation. We observe a sudden transition of $\eta_{<}^{2}$ at $\ell\sim\mathcal{O}(2\gamma/\sigma)$. In the regime $\ell<\mathcal{O}(2\gamma/\sigma)$, if separation is shorter than another critical value, the late time entanglement can survive and ameliorate with even smaller separation. This is a consequence of the interplay between the direct coupling and the indirect mutual influence between the oscillators. When $\ell>\mathcal{O}(2\gamma/\sigma)$ direct coupling $\sigma$ dominates, so entanglement in general improves with larger $\sigma$. On the other hand, when $\ell<\mathcal{O}(2\gamma/\sigma)$ the indirect, non-Markovian effect takes over the control of  the overall behavior.  Finally,  we note that the curves of $\eta_{<}^{2}$ sink down steeply while moving to smaller values of $\ell$ with larger $\sigma$.}\label{Fi:etavssigma}
\end{figure}

\subsection{$\varsigma>1$ : Direct coupling dominates}\label{S:gdngdkfgn2}
We first discuss the case $\varsigma>1$. This is the regime where the direct coupling dominates over the indirect mutual influence. The inter-oscillator coupling effectively correlated motions of both oscillators, so entanglement tends to survive over finite separation between them. Since there is no restriction about $\omega_{-}\ell$, we will discussion the critical parameters for two cases:  $\omega_{-}\ell>1$ and $\omega_{-}\ell<1$.

\subsubsection{$\omega_{-}\ell>1$}
When \textit{the inter-oscillator coupling is sufficiently strong},  that is $\varsigma>1$, the symplectic eigenvalue $\eta_{<}^{2}$, determined by $\langle\chi_{+}^{2}(\infty)\rangle\langle p_{-}^{2}(\infty)\rangle$, is given by
\begin{equation}\label{E:hrjsksa}
	\eta_{<}^{2}=\frac{g(\omega,\gamma,\sigma,\Lambda)}{\pi^{2}\Omega_{-}}\biggl[\frac{1}{\Omega_{+}}\,\cot^{-1}\frac{\gamma}{\Omega_{+}}+\frac{2\gamma}{(\Omega_{+}^{2}+\gamma^{2})^{2}}\frac{1}{\ell^{2}}-\frac{8\gamma^{2}}{(\Omega_{+}^{2}+\gamma^{2})^{3}}\frac{1}{\ell^{3}}+\mathcal{O}(\frac{1}{\ell^{4}})\biggr]\,,
\end{equation}
in the case $\omega_{\pm}\ell>1$, where $g(\omega,\gamma,\sigma,\Lambda)$ is positive and takes the form
\begin{equation}
	g(\omega,\gamma,\sigma,\Lambda)=\bigl(\Omega_{-}^{2}-\gamma^{2}\bigr)\cot^{-1}\frac{\gamma}{\Omega_{-}}+\gamma\Omega_{-}\ln\frac{\Lambda^{2}}{\Omega_{-}^{2}+\gamma^{2}}\,,
\end{equation}
with the resonance frequency $\Omega_{\pm}^{2}=\omega^{2}_{\pm}-\gamma^{2}$. The symplectic eigenvalue $\eta_{<}$ approaches a constant when the two oscillators are very far apart. If we write this asymptotic constant explicitly in terms of the coupling constants, we find
\begin{equation}
	\lim_{\ell\to\infty}\eta_{<}^{2}\simeq\frac{1}{4}\left[\frac{\omega^{2}-\sigma}{\omega^{2}+\sigma}\right]^{\frac{1}{2}}+\gamma\left[-\frac{\sqrt{\omega^{2}-\sigma}}{2\pi(\omega^{2}+\sigma)}+\frac{\sqrt{\omega^{2}+\sigma}}{2\pi(\omega^{2}+\sigma)}\left(\ln\frac{\Lambda^{2}}{\omega^{2}-\sigma}-1\right)\right]+\mathcal{O}(\gamma^{2})\,.
\end{equation}
This constant is smaller with larger $\sigma$ but grows with increasing $\gamma$ because the coefficient of $\gamma$ is always positive. It implies when the interaction between the oscillators and the bath is not negligibly small, we need sufficiently strong inter-oscillator coupling to overcome the vacuum fluctuations of the bath and to maintain entanglement between oscillators. For a given strength of the inter-oscillator coupling $\sigma$, the damping constant $\gamma$ cannot be greater than
\begin{equation}\label{E:dkrowow}
	\gamma<\dfrac{\frac{1}{4}\left(1-\sqrt{\dfrac{\omega^{2}-\sigma}{\omega^{2}+\sigma}}\right)}{\dfrac{\sqrt{\omega^{2}+\sigma}}{2\pi(\omega^{2}+\sigma)}\left(\ln\dfrac{\Lambda^{2}}{\omega^{2}-\sigma}-1\right)-\dfrac{\sqrt{\omega^{2}-\sigma}}{2\pi(\omega^{2}+\sigma)}}\,,
\end{equation}
otherwise the late time entanglement can not exist. For weaker inter-oscillator coupling $\sigma/\omega^{2}<1$, Eq.~\eqref{E:dkrowow} is approximately given by
\begin{equation}
	\frac{\gamma}{\omega}<\frac{\pi}{4(\ln\Lambda/\omega-1)}\frac{\sigma}{\omega^{2}}+\mathcal{O}(\frac{\sigma^{2}}{\omega^{4}})\,.
\end{equation}
This upper bound depends on the cutoff scale $\Lambda$, which originates from the momentum uncertainty. Thus the inter-oscillator and the oscillator-bath couplings play competing roles. Intuitively because the damping constant $\gamma=e^{2}/8\pi m$ is a measure  of how much the quantum (vacuum or thermal) fluctuations of the bath can possibly disrupt the correlations between the oscillators, so larger $\gamma$ tends to make the influence of the bath more destructive on the correlations between the oscillators.  On the other hand  inter-oscillator coupling is expected to increase the bond between the two oscillators.  Nonetheless, the mechanism of how entanglement is affected by this direct coupling is not as intuitive as one would have superficially assumed because when we look into the motion of the normal modes of the joint system, we note that it causes the opposite effects. More specifically the stronger inter-oscillator coupling will make the CoM mode oscillate faster, which in turn decreases (increases) its position (momentum) uncertainty, meanwhile it reduces (amplifies) the position (momentum) uncertainty of the relative mode. However from our qualitative discussions based on the effective frequencies, it is not difficult to see how these come into play regarding the existence of entanglement. In short, strong inter-oscillator and weaker bath-oscillator interaction benefit entanglement only for the case $\varsigma>1$.   

\begin{figure}
	\centering
    \scalebox{0.5}{\includegraphics{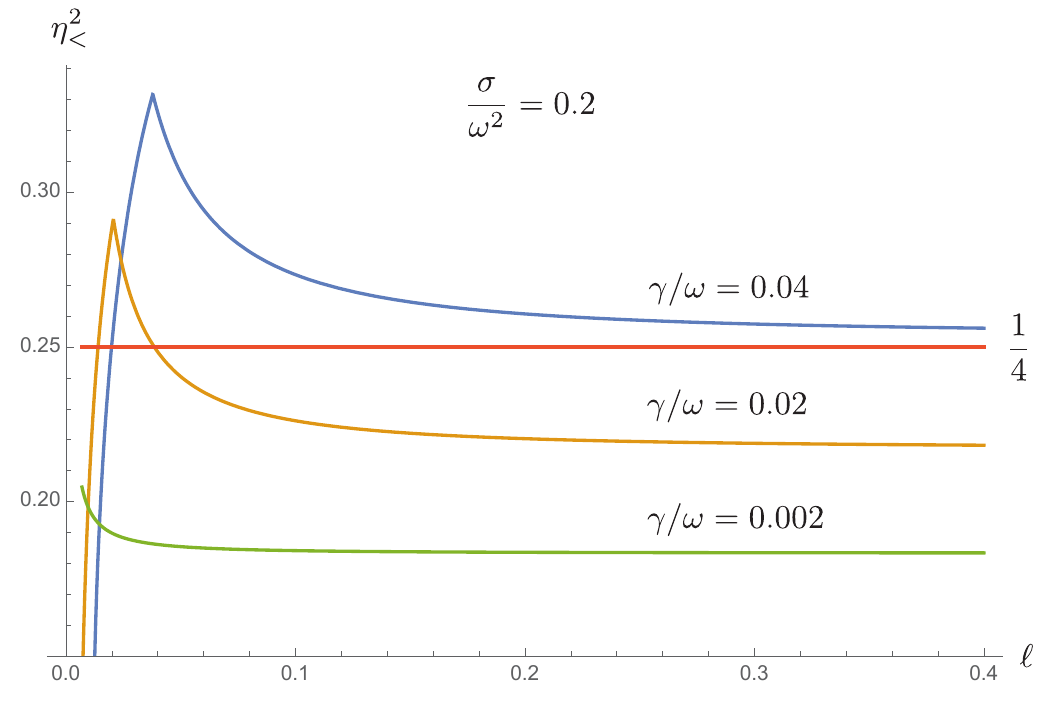}}
    \captionsetup{justification=raggedright}
    \caption{The curves of $\eta_{<}^{2}$ move upward and to the right with larger $\gamma$, so at a fixed separation this implies smaller values of $\gamma$ are in favor of entanglement . In the regime $\varsigma>\mathcal{O}(1)$ the critical separation increases with larger $\gamma$, implying that the late time entanglement is harder to sustain. On the other hand, for $\varsigma<\mathcal{O}(1)$ larger values of $\gamma$ in fact improves entanglement. Comparing this with Fig.~\ref{Fi:etavssigma}, we see the effects of $\sigma$ and $\gamma$ on $\eta_{<}^{2}$, that is, on entanglement, are totally opposite in the two regimes $\varsigma<\mathcal{O}(1)$ and  $\varsigma>\mathcal{O}(1)$ .}\label{Fi:etavsgamma}
\end{figure}

Since the leading term that depends on the separation $\ell$ in \eqref{E:hrjsksa} is positive, we see that as the separation $\ell$ decreases the value of $\eta_{<}^{2}$ increases as fast as $\ell^{-2}$ until $\eta_{<}^{2}$ reaches the value $1/4$, where entanglement is about to disappear. For weaker but sufficiently strong inter-oscillator coupling, the disentanglement may already happen at comparatively large separations. Let $\ell_{>}$ denote this critical separation,  we find
\begin{equation}\label{E:oqanpret}
	\ell_{>}=\left(\frac{4\beta}{1-4\alpha}\right)^{\frac{1}{2}}\,,
\end{equation}
where $\alpha$, $\beta$ are functions of $\omega$, $\gamma$ and $\sigma$,
\begin{align}
	\alpha&=\frac{\omega_{-}}{4\omega_{+}}+\frac{-\omega_{-}+\omega_{+}\bigl(\ln\dfrac{\Lambda^{2}}{\omega_{-}^{2}}-1\bigr)}{2\pi\omega_{+}^{2}}\,\gamma-\frac{\pi^{2}\bigl(3\omega_{+}^{2}-\omega_{-}^{2}\bigr)+8\omega_{+}\omega_{-}\bigl(\ln\dfrac{\Lambda^{2}}{\omega_{-}^{2}}-1\bigr)}{8\pi^{2}\omega_{+}^{3}\omega_{-}}\,\gamma^{2}+\mathcal{O}(\gamma^{3})\,,\\
	\beta&=\frac{\omega_{-}}{\pi\omega_{+}^{4}}\,\gamma+\frac{2\bigl(\ln\dfrac{\Lambda^{2}}{\omega_{-}^{2}}-1\bigr)}{\pi\omega_{+}^{4}}\,\gamma^{2}+\mathcal{O}(\gamma^{3})\,,
\end{align}
with $\omega_{\pm}^{2}=\omega^{2}\pm\sigma$. We will briefly highlight the derivation. In principle the solution of \eqref{E:hrjsksa} can be found exactly because it can be made into a third-order polynomial; however the solutions are too intricate to be of any practical use. On observing that the coefficient of the $\ell^{-3}$ term in \eqref{E:hrjsksa} is relatively small due to higher power in $\dfrac{\gamma}{\omega_{+}^{2}\ell}$, so we can drop the $\ell^{-3}$ term for the case $\omega_{+}\ell>1$ and write \eqref{E:hrjsksa} into the form
\begin{equation}\label{E:sqenfkd}
	\eta_{<}^{2}\simeq\alpha+\frac{\beta}{\ell^{2}}\simeq\frac{1}{4}\,,
\end{equation}
after we Taylor-expand \eqref{E:hrjsksa} in orders of the small parameter $\gamma/\omega_{\pm}$. Eq.~\eqref{E:sqenfkd} is equivalent to a second-order polynomial in $\ell$, and one of whose solutions is then \eqref{E:oqanpret}. At this step, we see the parameters $\alpha$, $\beta$ in \eqref{E:oqanpret} are still quite complicated and may attempt to further expand \eqref{E:oqanpret} in terms of small ratio $\gamma/\omega_{\pm}$. It turns out that it may not always work, especially for the case {when $\ell$ is sufficiently large, the saturated value of the curve $\eta_{<}^{2}$ happens to lie right below the horizontal like $1/4$ (please refer to Fig.~\ref{Fi:etavsgamma}). To meet this situation, the coupling constants $\gamma$ and $\sigma$ must be so chosen to make the value of $\eta_{<}^{2}$ for far-apart oscillators close to but smaller than} the critical value $1/4$, that is, 
\begin{equation}\label{E:kqqskjer}
	\lim_{\Omega_{+}\ell>1}\eta_{<}^{2}\precsim\frac{1}{4}\,.
\end{equation}
Since both $\beta$ and $(1-4\alpha)$ are already small, any minute variation may induce large error in approximation. On top of that, the $\mathcal{O}(1)$ and $\mathcal{O}(\gamma)$ terms in $(1-4\alpha)$ are comparable in magnitude but take on opposite signs. This makes $(1-4\alpha)$  at least roughly of the order $\mathcal{O}(\gamma^{2})$. It may appear odd how the $\mathcal{O}(1)$ and $\mathcal{O}(\gamma)$ terms in $(1-4\alpha)$ can be comparable even for small $\gamma$.  But recall that in order to (almost) satisfy \eqref{E:kqqskjer}, the coupling constants $\gamma$ and $\sigma$ must take on particular values. The consequence of that is the (almost) cancellation of the $\mathcal{O}(1)$ and $\mathcal{O}(\gamma)$ terms in $(1-4\alpha)$. This is not straightforwardly revealed via the Taylor expansion of small $\gamma$ and will yield erroneous results for the case described by \eqref{E:kqqskjer}. Thus \eqref{E:oqanpret} can give a satisfactory estimation of $\ell_{>}$ even $\Omega_{\pm}\ell_{>}$ is large, but it still does not work for the case $\Omega_{\pm}\ell_{>}\gg1$, where the contributions of the order higher than $\mathcal{O}(\gamma^{2})$ must be taken into consideration.

\subsubsection{$\omega_{-}\ell<1$}
As seen in Fig.~\ref{Fi:etavssigma} when the inter-oscillator coupling is stronger, disentanglement could happen at relatively small separation. Thus we use the short-distance approximation ($\omega_{\pm}\ell<1$) of $\eta_{<}^{2}$
\begin{align}
	\eta_{<}^{2}&=\biggl[\frac{\omega_{-}^{3}}{4\omega_{+}^{3}}-\frac{(\omega_{+}+\omega_{-})\omega_{-}^{2}}{\pi\omega_{+}^{3}}\frac{\gamma}{\omega_{+}}+\mathcal{O}(\frac{\gamma^{2}}{\omega_{+}^{2}})\biggr]\frac{\gamma^{2}}{\omega_{-}^{4}\ell^{2}}\notag\\
	&\qquad\qquad\qquad+\biggl[\frac{(\omega_{+}^{2}+\omega_{-}^{2})\omega_{-}}{4\omega_{+}^{3}}-\frac{(\omega_{+}+\omega_{-})(\omega_{+}^{2}+\omega_{-}^{2})-\omega_{+}\omega_{-}^{2}(\gamma_{\epsilon}+\ln\Lambda\ell)}{\pi\omega_{+}^{3}}\frac{\gamma}{\omega_{+}}+\mathcal{O}(\frac{\gamma^{2}}{\omega_{+}^{2}})\biggr]\frac{\gamma}{\omega_{-}^{2}\ell}\notag\\
	&\qquad\qquad\qquad\qquad\qquad\qquad+\biggl[\frac{\omega_{-}}{4\omega_{+}}-\frac{(\omega_{+}+\omega_{-})-\omega_{+}(\gamma_{\epsilon}+\ln\Lambda\ell)}{\pi\omega_{+}}\frac{\gamma}{\omega_{+}}+\mathcal{O}(\frac{\gamma^{2}}{\omega_{+}^{2}})\biggr]+\mathcal{O}(\frac{\ell}{\gamma\omega_{-}^{2}})\,,\notag\\
	&=\Bigl[a_{0}+a_{1}\,\frac{\gamma}{\omega_{+}}+\cdots\Bigr]\frac{\gamma^{2}}{\omega_{-}^{4}\ell^{2}}+\Bigl[b_{0}+b_{1}\,\frac{\gamma}{\omega_{+}}+\cdots\biggr]\frac{\gamma}{\omega_{-}^{2}\ell}+\biggl[c_{0}+c_{1}\,\frac{\gamma}{\omega_{+}}+\cdots\biggr]+\mathcal{O}(\frac{\ell}{\gamma\omega_{-}^{2}})\,,\label{E:qsnkfd}
\end{align}
to find the critical separation $\ell_{>}$. In \eqref{E:qsnkfd} the shorthand notations are
\begin{align*}
	a_{0}&=\frac{\omega_{-}^{3}}{4\omega_{+}^{3}}\,,&b_{0}&=\frac{(\omega_{+}^{2}+\omega_{-}^{2})\omega_{-}}{4\omega_{+}^{3}}\,,&b_{1}&=-\frac{(\omega_{+}+\omega_{-})(\omega_{+}^{2}+\omega_{-}^{2})-\omega_{+}\omega_{-}^{2}(\gamma_{\epsilon}+\ln\Lambda\ell)}{\pi\omega_{+}^{3}}\,,\\
	a_{1}&=-\frac{(\omega_{+}+\omega_{-})\omega_{-}^{2}}{\pi\omega_{+}^{3}}\,,&c_{0}&=\frac{\omega_{-}}{4\omega_{+}}\,, &c_{1}&=-\frac{(\omega_{+}+\omega_{-})-\omega_{+}(\gamma_{\epsilon}+\ln\Lambda\ell)}{\pi\omega_{+}}\,.
\end{align*}
By means of the criterion $\eta_{<}^{2}=1/4$, the critical separation $\ell_{>}$ is determined approximately by
\begin{equation}\label{E:lkiqnkzaq}
	\ell_{>}=\ell_{>}^{(0)}+\frac{a_{1}\left(\dfrac{\omega_{-}^{2}\ell_{>}^{(0)}}{\gamma}\right)+b_{1}\left(\dfrac{\omega_{-}^{2}\ell_{>}^{(0)}}{\gamma}\right)^{2}+c_{1}\left(\dfrac{\omega_{-}^{2}\ell_{>}^{(0)}}{\gamma}\right)^{3}}{2a_{0}+b_{0}\left(\dfrac{\omega_{-}^{2}\ell_{>}^{(0)}}{\gamma}\right)}\frac{\gamma}{\omega_{+}}+\mathcal{O}(\frac{\gamma^{2}}{\omega_{+}^{2}})\,,
\end{equation}
where $\ell_{>}^{(0)}$ is the leading contribution in the critical separation
\begin{equation}
	\ell_{>}^{(0)}=\frac{2\gamma}{-(\omega_{+}^{2}+\omega_{-}^{2})+[(\omega_{+}^{2}-\omega_{-}^{2})^{2}+4\omega_{+}^{3}\omega_{-}]^{1/2}}\,,
\end{equation}
To arrive at \eqref{E:lkiqnkzaq} we have made the assumption $\gamma\ll\omega_{\pm}$ to simplify the result and recall that $\omega_{\pm}^{2}=\omega^{2}\pm\sigma$.

\subsubsection{Comments}
In this regime the choice of $\sigma$ is limited by two constraints: $\varsigma>1$ and $2\gamma<\omega_{-}^{2}\ell$. They imply that 
\begin{equation}\label{E:ekjkssw}
	\frac{2\gamma}{\ell}<\sigma<\omega^{2}-\frac{2\gamma}{\ell}\,.
\end{equation}
For small values of $\sigma$, the critical length tends to be large and \eqref{E:oqanpret} gives a better approximation, while large $\sigma$ implies a small critical separation so \eqref{E:lkiqnkzaq} is more suitable. Within the range \eqref{E:ekjkssw}, the critical separation $\ell_{c}$ decreases with growing inter-oscillator coupling strength. The physical interpretation regarding the separation of the oscillators is much less straightforward due to the fact that non-zero separation picks up the correlations existent in the bath between two spatial locations and is the cause of non-Markovianity in the propagation of mutual influences. Since the non-Markovian influence of one oscillator will propagate by means of the intermediate bath in the form of the retarded radiation to the other oscillator, thus inevitable introducing a disparate phase and further degrading coherence between them. However, this non-Markovian effect is greatly reduced by the separation. That may offer a physical understanding why in this regime entanglement can possibly improve with increasing separation between oscillators.

\subsection{$\varsigma\sim\mathcal{O}(1)$}
When we further decrease the separation between oscillators, the joint system eventually evolves into a separable state.  Direct coupling is now not strong enough to sustain the late-time entanglement. We will gradually come to another critical separation $\ell_{\times}\sim2\gamma/\sigma$, where the crossover between the values of $\langle\chi_{+}^{2}(\infty)\rangle\langle p_{-}^{2}(\infty)\rangle$ and $\langle\chi_{-}^{2}(\infty)\rangle\langle p_{+}^{2}(\infty)\rangle$ happens. Once the separation is smaller than this critical value, the symplectic eigenvalue will be represented by $\langle\chi_{-}^{2}(\infty)\rangle\langle p_{+}^{2}(\infty)\rangle$ instead of $\langle\chi_{+}^{2}(\infty)\rangle\langle p_{-}^{2}(\infty)\rangle$ because now the former is the smaller among the two.

\subsection{$\varsigma<1$: non-Markovian interaction dominates}\label{S:gdngdkfgn4}
In this regime the non-Markovian effects is more important and the effects of direct coupling becomes subdominant, so we expect that the results should be similar to the earlier works~\cite{LH09, Zell09}, where direct coupling is absent. The stability criterion implies that 
\begin{align}
	\frac{2\gamma}{\omega_{-}^{2}}&<\ell<\frac{2\gamma}{\sigma}\,,&&\text{and}&\sigma&<\frac{\omega^{2}}{2}\,.
\end{align}
Thus when the inter-oscillator coupling is very weak, i.e., $\sigma/\omega\ll1$, we find $\omega\ell$ can be greater than one. On the other hand, if $\sigma/\omega^{2}\leq1/2$, then the value of $\omega\ell$ is very small, at most about the order $\mathcal{O}(\gamma/\omega^{2})$.

\subsubsection{$\omega_{-}\ell\gg1$}
First considering the case $\sigma/\omega^{2}\ll1$, we use the large separation approximation, that is, $\omega_{-}\ell\gg1$, to express $\eta_{<}^{2}$,
\begin{align}
	\eta_{<}^{2}&=\biggl\{\frac{\omega_{+}}{4\omega_{-}}+\Bigl[-\frac{\omega_{+}}{2\pi\omega_{-}}+\frac{1}{2\pi}\Bigl(\ln\frac{\Lambda^{2}}{\omega_{+}^{2}}-1\Bigr)\Bigr]\frac{\gamma}{\omega_{-}}-\Bigl[\frac{1}{\pi^{2}}\Bigl(\ln\frac{\Lambda^{2}}{\omega_{+}^{2}}-1\Bigr)+\frac{3\omega_{-}^{2}-\omega_{+}^{2}}{8\omega_{+}\omega_{-}}\Bigr]\frac{\gamma^{2}}{\omega_{-}^{2}}+\cdots\biggr\}\notag\\
	&\qquad\qquad\qquad\qquad\qquad+\biggl\{-\frac{\omega_{+}}{\pi\omega_{-}}\frac{\gamma}{\omega_{-}}+\frac{2}{\pi^{2}}\Bigl(\ln\frac{\Lambda^{2}}{\omega_{+}^{2}}-1\Bigr)\frac{\gamma^{2}}{\omega_{-}^{2}}+\cdots\biggr\}\frac{1}{\omega_{-}^{2}\ell^{2}}+\cdots\,.
\end{align}
Again it shows that for $\varsigma<1$ but $\omega_{-}\ell\gg1$, the oscillators remain separable until the distance between them is shortened to $\ell=\mathcal{O}(\omega_{-}^{-1})$, where $\eta_{<}^{2}$ starts decreasing rapidly. Apparently it is not suitable for identifying the critical separation. Hence we need an approximation of $\eta_{<}^{2}$ that works better in the regime $\omega_{-}\ell<1$.

\subsubsection{$\omega_{-}\ell\ll1$}
We find the small-separation approximation of $\eta_{<}^{2}$ is roughly given by
\begin{align}
	\eta_{<}^{2}&\simeq\biggl[\frac{\omega_{-}}{4\omega_{+}}-\frac{\omega_{+}+\omega_{-}}{\pi\omega_{+}}\frac{\gamma}{\omega_{+}}+\cdots\biggr]\frac{\gamma^{2}}{\omega_{-}^{4}\ell^{2}}+\biggl[-\frac{\omega_{+}^{2}+\omega_{-}^{2}}{4\omega_{+}\omega_{-}}+\frac{(\omega_{+}^{3}+\omega_{-}^{3})+\omega_{+}^{2}\omega_{-}(\gamma_{\epsilon}+\ln\dfrac{\omega_{+}^{2}\ell}{\Lambda})}{\pi\omega_{+}\omega_{-}^{2}}\frac{\gamma}{\omega_{+}}+\cdots\biggr]\frac{\gamma}{\omega_{-}^{2}\ell}\notag\\
	&\qquad\qquad\qquad\qquad\qquad\qquad+\biggl[\frac{\omega_{+}}{4\omega_{-}}-\frac{\omega_{+}(\gamma_{\epsilon}+\ln\dfrac{\omega_{+}^{2}\ell}{\Lambda})}{\pi\omega_{-}}\frac{\gamma}{\omega_{+}}+\cdots\biggr]+\mathcal{O}(\ell)\,.\label{E:qwqoirez}
\end{align}
The leading term of the order $\dfrac{\gamma}{\omega_{-}^{2}\ell}$ is negative. In addition, in this regime $\sigma<2\gamma/\ell$, the stability criterion implies 
\begin{equation}
	\frac{2\gamma}{\omega^{2}\ell}<\frac{2\gamma}{\omega_{-}^{2}\ell}<1\,,
\end{equation}
if we recall $\omega_{\pm}^{2}=\omega^{2}-\gamma^{2}$. Therefore as far as the leading terms in each pair of brackets in \eqref{E:qwqoirez} are concerned, we find
\begin{equation}
	\frac{1}{4\omega_{+}\omega_{-}^{3}}\frac{\gamma^{2}}{\ell^{2}}-\frac{\omega_{+}^{2}+\omega_{-}^{2}}{4\omega_{+}\omega_{-}^{3}}\frac{\gamma}{\ell}+\cdots\simeq\frac{\omega_{+}^{2}+\omega_{-}^{2}}{4\omega_{+}\omega_{-}^{3}}\frac{\gamma}{\ell}\biggl[\frac{\gamma}{2\omega^{2}\ell}-1\biggr]+\cdots<0\,,
\end{equation}
so $\eta_{<}^{2}$ decreases with smaller separation $\ell$ until again it crosses the value $1/4$. This means that entanglement is likely to exist and furthermore is improved for shorter separation. 
\begin{figure}
	\centering
    \scalebox{0.6}{\includegraphics{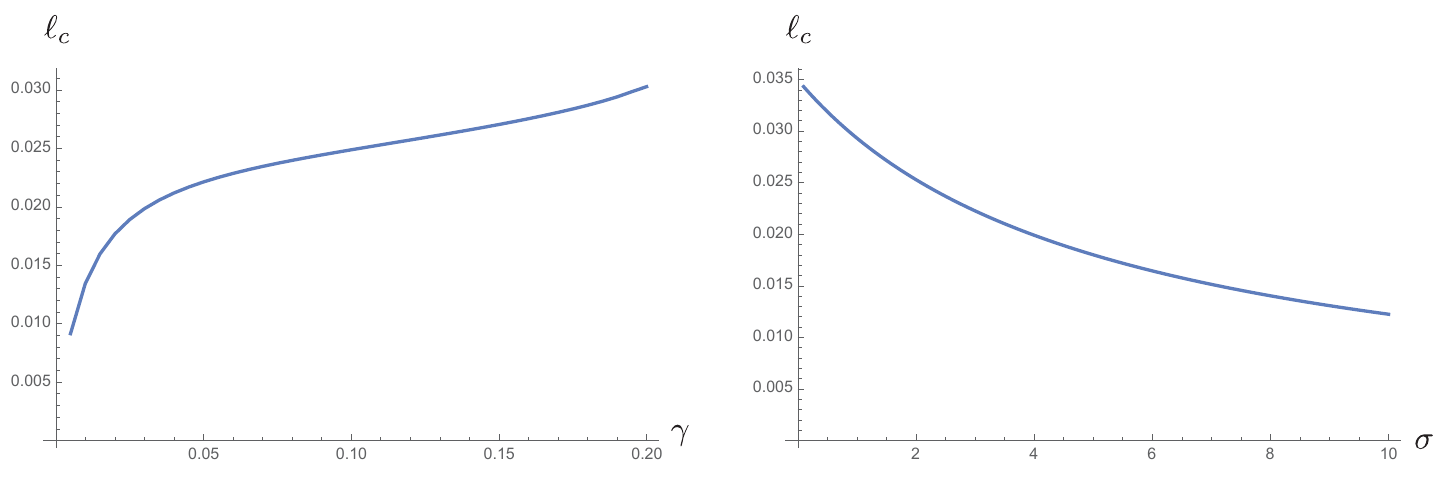}}
    \captionsetup{justification=raggedright}
    \caption{By numerically solving for $\eta_{<}^{2}=1/4$, we generate the relation between the critical separation $\ell_{<}$ with respect to $\gamma$ and $\sigma$ for the case described by \eqref{E:dislwsd} in the regime $\varsigma<1$.  Late time entanglement indeed improves with larger $\gamma$ and smaller $\sigma$. In this regime, the indirect non-Markovian mutual influence dominates over the effects of direct coupling $\sigma$ between the oscillators.  Such history-dependent effects make the physical interpretation less straightforward. In the plot we choose the oscillating frequency $\omega=5$ and the cutoff scale $10000$.}\label{Fi:crtisep}
\end{figure}

Compared with \eqref{E:qsnkfd} where the cutoff-dependent term takes the form $\ln\Lambda\ell$, the counterpart in \eqref{E:qwqoirez} has quite a distinct form, $\ln\dfrac{\omega_{+}^{2}}{\Lambda}\,\ell$. Thus the logarithmic cutoff dependent terms in \eqref{E:qwqoirez} weigh much more than those in \eqref{E:qsnkfd} unless $\gamma$ is vanishingly tiny, and this makes the application of the iteration scheme in \eqref{E:qwqoirez} much trickier. On account of this difficulty, we observe in \eqref{E:qwqoirez} that the contribution in the first pair of brackets on the righthand side can be negligible compared with the rest. If we ignore it, the remaining expressions of \eqref{E:qwqoirez} can be used to formally solve $\eta_{<}^{2}=1/4$, leading to
\begin{equation}\label{E:dislwsd}
	\ell_{<}=\frac{\biggl[\dfrac{\omega_{+}^{2}+\omega_{-}^{2}}{4\omega_{+}\omega_{-}}-\dfrac{(\omega_{+}^{3}+\omega_{-}^{3})+\omega_{+}^{2}\omega_{-}(\gamma_{\epsilon}+\ln\dfrac{\omega_{+}^{2}\ell_{<}}{\Lambda})}{\pi\omega_{+}\omega_{-}^{2}}\dfrac{\gamma}{\omega_{+}}\biggr]\dfrac{\gamma}{\omega_{-}^{2}}}{\dfrac{\omega_{+}-\omega_{-}}{4\omega_{-}}-\dfrac{\omega_{+}(\gamma_{\epsilon}+\ln\dfrac{\omega_{+}^{2}\ell_{<}}{\Lambda})}{\pi\omega_{-}}\dfrac{\gamma}{\omega_{+}}}\,,
\end{equation}
Repeated iterations in $\ell_{<}$ on the righthand side of \eqref{E:dislwsd} rapidly produce better improvement on the critical separation $\ell_{<}$, although the result can become very complicated in a couple of iterations.

This approximation works best for smaller values of $\sigma$ because the ratio $\dfrac{2\gamma}{\omega_{-}^{2}\ell_{<}}$ will remain small. From the involved expression of \eqref{E:dislwsd} we can still identify the facts that \textit{the critical separation grows with increasing $\gamma$ and decreasing $\sigma$. }In Fig.~\ref{Fi:crtisep} the results, generated numerically, support this trend. On the contrary, \textit{stronger inter-oscillator coupling will render this critical separation even shorter}, but in the end we will come to a situation that either the approximation breaks down or the instability occurs.

\begin{figure}
	\centering
    \scalebox{0.5}{\includegraphics{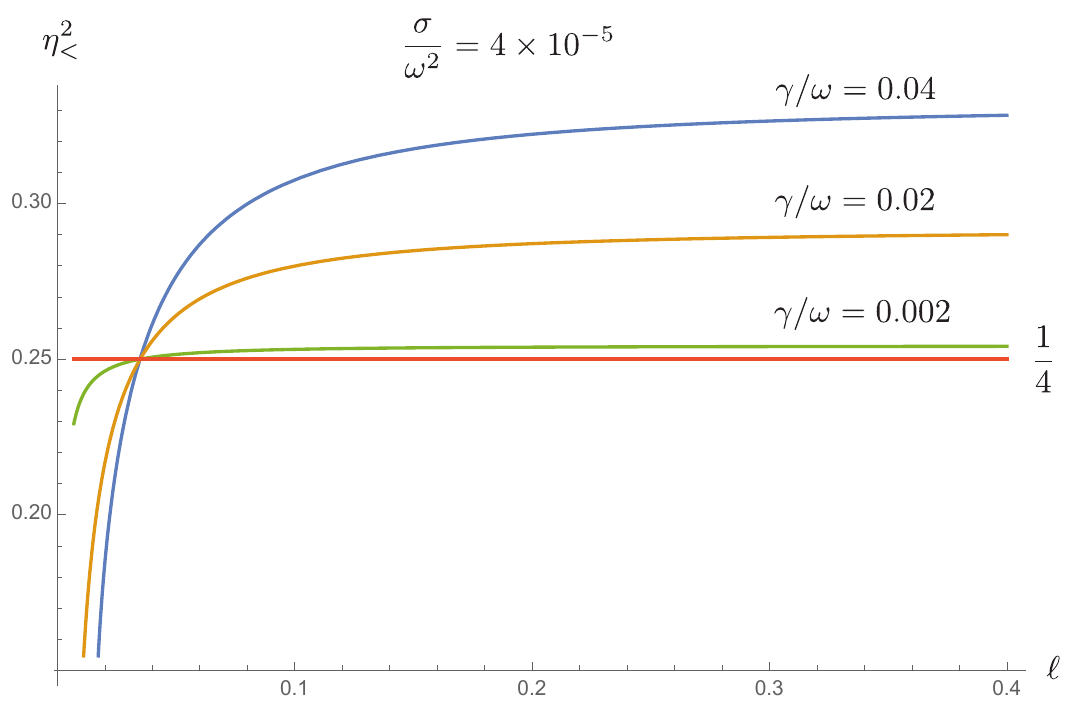}}
    \captionsetup{justification=raggedright}
    \caption{When the inter-oscillator coupling is vanishingly small, the critical separations are almost independent of the damping constant $\gamma$. Note that since the inter-oscillator coupling is too weak, there is only one critical separation for each curve. It is impossible to sustain entanglement at large separation in this case.}\label{Fi:etavsgammasml}
\end{figure}

In the limit $\sigma/\omega^{2}\ll\gamma/\omega$, that is, with negligible direct coupling between the oscillators, the critical separation $\ell_{<}$ approaches  a value with very mild dependence on $\gamma$. If the ratio $\gamma/\omega$ is much smaller than one, then this value essentially becomes a constant,
\begin{align}\label{E:mxnewa} 
	\ell_{<}&=\frac{\pi}{2\omega\,(\ln\dfrac{\Lambda}{\omega^{2}\ell_{<}}-\gamma_{\epsilon})}\,,&&\text{for}\qquad\frac{\sigma}{\omega^{2}}\ll\frac{\gamma}{\omega}\ll1\,,
\end{align}
as shown in Fig.~\ref{Fi:etavsgammasml}. Since this is the iterative expression, if we truncate it to the first order, we can substitute $\ell_{<}$ in the logarithm on the righthand side by $\dfrac{\pi}{2\omega}$. 

So far, we note that when we look for the critical separation, we repeatedly come across solving a transcendental function of the form
\begin{equation} 
	\frac{c_{2}}{\ell}+c_{1}\ln\ell+c_{0}=0\,.
\end{equation}
Some of its solutions, as shown previously, can be found by iterative substitutions, but in fact it can be expressed in terms of a special function, the product logarithm or Lambert $\mathcal{W}$-function $\mathcal{W}(z)$, which is the principal solution for $w$ in $z=w\,e^{w}$. Hence for example, Eq.~\eqref{E:mxnewa} can be written as
\begin{equation}
	\ell_{c}=-\frac{\pi}{2\omega}\left[\mathcal{W}(-1,-\frac{\pi\omega}{2\Lambda}\,e^{\gamma_{\epsilon}})\right]^{-1}\,,
\end{equation}
where $\mathcal{W}(k,z)$ is the $k$th solution to $z=w\,e^{w}$ other than the principal one. In addition, the trails of the cutoff scale are dotted everywhere and they cannot always been argued away. As in the case of \eqref{E:mxnewa}, the effect of the cutoff scale persists and is independent of the damping constant $\gamma$ even though $\gamma$ is already tiny.

Thus we see, \textit{for sufficiently weak inter-oscillator coupling, when the separation between the oscillators is less than this critical value, the entanglement between them can survive at late time}. If we further reduce the separation, we will at last come to the situation that the dynamics of the joint system becomes unstable.

\section{Summary of Results}

Here we summarize our analytical results and numerical supports as shown in the figures. We first note that different normal modes have different effective relaxation time scales, and these time scales also depend on the separation between oscillators. When the mutual separation is small $\omega_{\pm}\ell\ll1$, the motion of the slow mode relaxes much slower than that of the fast mode. Thus the late-time results will not be valid unless the motion of all normal modes are relaxed. Within the relaxation time, we assume the mutual influence between the oscillators has been exchanged for sufficiently many times. Then we find that except for some extreme cases, typically speaking, in the regime $\varsigma>1$, the late time entanglement improves with 1) stronger inter-oscillator coupling, 2) weaker bath-oscillator coupling, or 3) longer oscillator separation. The critical separation at which disentanglement occurs thus reduces with larger value of $\sigma$ but smaller value of $\gamma$, as can be seen in Figs.~\ref{Fi:etavssigma} and~\ref{Fi:etavsgamma}. In addition, within this regime, the direct coupling between oscillators dominates over the indirect, non-Markovian interaction. The latter effect depends on oscillators' histories and is manifested as the retarded radiation mediated by the environment. Thus the magnitude and the phase of this non-Markovian effect will depend on the separation. Roughly speaking at large separation, we see from the Fourier or Laplace transforms of the normal-mode equations of motion~\eqref{E:xmsdhs} that the retarded term tends to introduce the phase difference from the local terms for the contribution of each environment mode due to the delay by $\ell$. This out-of-phase effect is factored out by larger oscillator separation, so it more or less explains the improvement of entanglement with increase separation in this regime.

In the neighborhood $\varsigma\sim\mathcal{O}(1)$, comparison of the behavior of $\eta_{<}^{2}$ between various values of $\gamma$ or $\sigma$ is difficult to make because that is the region where the transition of $\eta_{<}^{2}$ happens. For example, in Fig.~\ref{Fi:etavssigma} when $\ell\simeq0.03$, the curve of $\eta_{<}^{2}$ for $\sigma/\omega^{2}=0.2$ is described by $\langle\chi_{-}^{2}(\infty)\rangle\langle p_{+}^{2}(\infty)\rangle$, but the curve that corresponds to $\sigma/\omega^{2}=0.4$ is still given by $\langle\chi_{+}^{2}(\infty)\rangle\langle p_{-}^{2}(\infty)\rangle$. Thus we don't see the comparison between them particularly meaningful.

When $\varsigma<1$, we find the entanglement tends to improve with 1) weaker inter-oscillator coupling, 2) stronger bath-oscillator interaction, or 3) shorter separation. In other words, stronger bath-oscillator interaction, or weak inter-oscillator coupling is preferable for the late time entanglement so that it can still be sustained at larger mutual separation. This conclusion can be confirmed in Fig.~\ref{Fi:crtisep}, where the critical separation is numerically found by solving $\eta_{<}^{2}=1/4$ for different values of $\gamma$ and $\sigma$.  These results seem less intuitive because in this regime the non-Markovian effect is much more prominent than the direct coupling effect. Except for the case that $\sigma$ is vanishingly small, the typical values of $\ell$ satisfy $\omega_{\pm}\ell<1$. This implies the delay introduces only a nominal phase difference between the retarded term and local terms for the contribution from each environment mode. This subtle difference can be very sensitive to the values of $\sigma$ and $\gamma$, and then is readily amplified by the tiny value of the mutual separation, as can be seen in~\eqref{E:xmsdhs}. This points out the elusive aspect of the non-Markovian phenomena.

However, the effective oscillating frequency $W_{\pm}$, introduced earlier, can be a very attractive tool in deciphering the non-Markovian effects. Recall that the corresponding effective frequency $W_{\pm}$ is defined by \eqref{E:xegsd}
\begin{equation*}
	W_{\pm}^{2}=\omega^{2}\pm\sigma\mp\frac{2\gamma}{\ell}\,,
\end{equation*}
where the non-Markovian effect is already encapsulated in the expression $2\gamma/\ell$. We see that this non-Markovian effect tends to lower (raise) the oscillating frequency of the CoM (relative) mode. The reduction (enhancement) increases with larger bath-oscillator coupling $\gamma$ and shorter separation $\ell$. This can be understood from the equations of motion of the normal modes 
\begin{equation*}
	\ddot{\chi_{\pm}}+2\gamma\,\dot{\chi}_{\pm}(s)+\omega^{2}_{\pm}\,\chi_{\pm}(s)=\pm2\gamma\,\frac{\theta(s-\ell)}{\ell}\,\chi_{\pm}(s-\ell)+\frac{1}{m}\,\xi_{\pm}(s)\,.
\end{equation*}
We see on the righthand side the delayed term in fact is the retarded Coulomb-like force. For the CoM mode, it exerts a repulsive force, which  counteracts with the restoring force from the harmonic potential of the CoM mode, effectively leading to slowdown of the CoM-mode oscillation. Likewise, this retarded Coulomb force is attractive for the relative mode, and thus works hand in hand with its restoring force, causing a more rapid  oscillation. The effect of the direct coupling can also be easily understood by means of this effective interpretation in the same fashion. Earlier on we have argued that when $\varsigma>\mathcal{O}(1)$, we have $W_{+}>W_{-}$ and the value of $\eta_{<}^{2}$ is roughly given by
\begin{equation}
	\eta_{<}^{2}\sim\mathcal{O}(\frac{W_{-}}{W_{+}})=\mathcal{O}(\,\biggl[\frac{\omega^{2}-\sigma+\dfrac{2\gamma}{\ell}}{\omega^{2}+\sigma-\dfrac{2\gamma}{\ell}}\biggr]^{\frac{1}{2}}\,)\,,
\end{equation}
if we explicitly spell out the dependence of $W_{\pm}$ on $\gamma$, $\sigma$, $\omega$ and $\ell$. T\textit{his clearly explains that the entanglement will improve with larger inter-oscillator coupling, weak bath-oscillator interaction, and longer separation between oscillators.} On the other hand, when $\varsigma<\mathcal{O}(1)$, we instead have $W_{-}>W_{+}$. In this case the value of $\eta_{<}^{2}$ takes a different form
\begin{equation}
	\eta_{<}^{2}\sim\mathcal{O}(\frac{W_{+}}{W_{-}})=\mathcal{O}(\,\biggl[\frac{\omega^{2}+\sigma-\dfrac{2\gamma}{\ell}}{\omega^{2}-\sigma+\dfrac{2\gamma}{\ell}}\biggr]^{\frac{1}{2}}\,)\,.
\end{equation}
It also nicely explicates why \textit{the interaction and the mutual separation play totally opposite roles in this case, }in comparison with their effects for the situation $\varsigma>\mathcal{O}(1)$.

In this paper we have  shown that  spatial dependence and direct coupling of two oscillators in a common zero-temperature quantum field environment engenders very rich features in their entanglement dynamics arising from the interplay between the different scales involved.
In a sequel paper  \cite{HHQEnt1} we will consider the entanglement between two coupled oscillators in a common finite temperature bath, and address the question whether and under what conditions entanglement can be sustained at high temperatures, as some authors alluded to recently \cite{hotEnt}. \\

\noindent {\bf Acknowledgment}  JTH thanks Shi-Yuin Lin for valuable discussions.  BLH enjoyed the hospitality of the Center for Field Theory and Particle Physics at Fudan University, China in January 2015 while engaging in this collaborative work.  \\

\appendix

\section{evaluation of $\langle\chi_{\pm}^{2}(t)\rangle$ and $\langle p _{\pm}^{2}(t)\rangle$ at zero temperature}\label{E:nvndfkse}
Here we first evaluate $\langle\chi_{\pm}^{2}(t)\rangle$. In general, at late time the expectation values of $\displaystyle\lim_{t\to\infty}\langle\chi^{2}_{\pm}(t)\rangle$ and $\displaystyle\lim_{t\to\infty}\langle p_{\pm}^{2}(t)\rangle$ have be shown respectively to be
\begin{align}
	\langle\chi^{2}_{\pm}(\infty)\rangle&=\frac{2c_{\pm}}{m}\operatorname{Im}\int_{0}^{\infty}\!\frac{d\kappa}{2\pi}\;\overline{d}_{2}^{(\pm)}(\kappa)\,\coth\frac{\beta\kappa}{2}\,,\\
	\langle p^{2}_{\pm}(\infty)\rangle&=2mc_{\pm}\operatorname{Im}\int_{0}^{\infty}\!\frac{d\kappa}{2\pi}\;\kappa^{2}\,\overline{d}_{2}^{(\pm)}(\kappa)\,\coth\frac{\beta\kappa}{2}\,,
\end{align}
for the CoM and the relative modes. However, due to the presence of the hypercotangent function and non-Markovianity inherent in $\overline{d}_{2}^{(\pm)}(\kappa)$, there is no closed form of the integral for arbitrary values of the parameters at hand. Hence we will evaluate $\langle\chi_{\pm}^{2}(\infty)\rangle$ and $\langle p_{\pm}^{2}(\infty)\rangle$ in the zero, the low- and the high-temperature limit, where the factor $\coth\dfrac{\beta\kappa}{2}$ is approximately given by
\begin{equation}
	\coth\frac{\beta\kappa}{2}=\begin{cases}
									1+2e^{-\beta\kappa}\,,&\beta\kappa\gg1\,,\\
									\dfrac{2}{\beta\kappa}+\dfrac{\beta\kappa}{6}+\cdots\,,&\beta\kappa\ll1\,.								
							   \end{cases}
\end{equation}
In addition, we will assume the coupling between the oscillator and the environment satisfies the condition $2\gamma<\omega^{2}\ell$ such that we can treat the non-Markovian term as small perturbation.

For the current article, we only discuss the zero temperature case, where $\coth\dfrac{\beta\kappa}{2}$ takes the value of unity, and we Taylor-expand $\overline{d}_{2}^{(\pm)}(\kappa)$ in terms of the small parameter $\dfrac{2\gamma}{\omega^{2}_{\pm}\ell}$
\begin{align}
	\overline{d}_{2}^{(\pm)}(\kappa)&=\frac{1}{-\kappa^{2}-i\,2\gamma\kappa+\omega_{\pm}^{2}}-\frac{1}{\omega_{\pm}}\frac{\partial}{\partial\omega_{\pm}}\biggl[\frac{\pm\dfrac{\gamma}{\ell}\,e^{i\,\kappa\ell}}{-\kappa^{2}-i\,2\gamma\kappa+\omega_{\pm}^{2}}\biggr]+\cdots
\end{align}
where $\overline{d}_{2}^{(\pm)}(\kappa)$ is obtained if we substitute $s=-i\,\kappa$ into $\tilde{d}_{2}^{(\pm)}(s)$ in \eqref{E:qjskerw}, that is, $\overline{d}_{2}^{(\pm)}(\kappa)=\tilde{d}_{2}^{(\pm)}(-i\,\kappa)$. We observe that since in general $\omega_{\pm}$ are different, we will restrict ourselves to the tighter constraint among the condition $\dfrac{2\gamma}{\omega^{2}_{\pm}\ell}$.

The expectation value $\langle\chi_{\pm}^{2}(\infty)\rangle$ in the zero temperature limit is given by
\begin{align}
	\langle\chi_{\pm}^{2}(\infty)\rangle&=\frac{2c_{\pm}}{m}\operatorname{Im}\int_{0}^{\infty}\!\frac{d\kappa}{2\pi}\;\biggl\{\frac{1}{-\kappa^{2}-i\,2\gamma\kappa+\omega_{\pm}^{2}}-\frac{1}{\omega_{\pm}}\frac{\partial}{\partial\omega_{\pm}}\biggl[\frac{\pm\dfrac{\gamma}{\ell}\,e^{i\,\kappa\ell}}{-\kappa^{2}-i\,2\gamma\kappa+\omega_{\pm}^{2}}\biggr]\biggr\}\,,
\end{align}
with $c_{+}=1/2$ for the CoM mode and $c_{-}=2$ for the relative mode. We begin with the calculation of the integral
\begin{align}
	\mathfrak{I}_{1}^{(\pm)}=\int_{0}^{\infty}\!d\kappa\;\frac{1}{-\kappa^{2}-i\,2\gamma\kappa+\omega_{\pm}^{2}}=\frac{i}{\Omega_{\pm}}\,\cot^{-1}\frac{\gamma}{\Omega_{\pm}}\,,
\end{align}
where $\Omega_{\pm}=\sqrt{\omega_{\pm}^{2}-\gamma^{2}}$ is the resonance frequency for the CoM/relative mode. Next the integral $\mathfrak{I}_{2}^{(\pm)}$ that accounts for the non-Markovian contribution can be expressed as differentiation of a simpler integral
\begin{equation}
	\mathfrak{I}_{2}^{(\pm)}=-\frac{1}{\omega_{\pm}}\frac{\partial}{\partial\omega_{\pm}}\int_{0}^{\infty}\!d\kappa\;\frac{\pm\dfrac{\gamma}{\ell}\,e^{i\,\kappa\ell}}{-\kappa^{2}-i\,2\gamma\kappa+\omega_{\pm}^{2}}\,,
\end{equation}
so we evaluate that integral first
\begin{align}
	\int_{0}^{\infty}\!d\kappa\;\frac{\dfrac{\gamma}{\ell}\,e^{i\,\kappa\ell}}{-\kappa^{2}-i\,2\gamma\kappa+\omega_{\pm}^{2}}&=i\,\frac{\gamma}{2\Omega_{\pm}\ell}\biggl[e^{+i\,(\Omega_{\pm}-i\,\gamma)\ell}\Bigl(\pi-i\,\operatorname{Ei}[-i\,(\Omega_{\pm}-i\,\gamma)\ell]\Bigr)\biggr.\notag\\
	&\qquad\qquad+\biggl.e^{-i\,(\Omega_{\pm}+i\,\gamma)\ell}\Bigl(\pi+i\,\operatorname{Ei}[+i\,(\Omega_{\pm}+i\,\gamma)\ell]\Bigr)\biggr]\,.
\end{align}
Thus we have $\mathfrak{I}_{2}^{(\pm)}$ given by
\begin{align}
	\mathfrak{I}_{2}^{(\pm)}=-i\,\frac{\gamma^{2}}{\Omega_{\pm}^{2}(\Omega_{\pm}^{2}+\gamma^{2})\ell}&+i\,\bigl(1-i\,\Omega_{\pm}\ell\bigr)\,\frac{\gamma}{2\Omega_{\pm}^{3}\ell}\,e^{+i\,(\Omega_{\pm}-i\,\gamma)\ell}\Bigl(\pi-i\,\operatorname{Ei}[-i\,(\Omega_{\pm}-i\,\gamma)\ell]\Bigr)\notag\\
	&+i\,\bigl(1+i\,\Omega_{\pm}\ell\bigr)\,\frac{\gamma}{2\Omega_{\pm}^{3}\ell}\,e^{-i\,(\Omega_{\pm}+i\,\gamma)\ell}\Bigl(\pi+i\,\operatorname{Ei}[+i\,(\Omega_{\pm}+i\,\gamma)\ell]\Bigr)\,.
\end{align}
We see that both $\mathfrak{I}_{1}^{(\pm)}$ and $\mathfrak{J}_{2}^{(\pm)}$ are pure imaginary. The exponential integral function $\operatorname{Ei}(z)$ is defined according to
\begin{equation}
	\operatorname{Ei}(z)=-\int^{\infty}_{-z}\!ds\;\frac{e^{-s}}{s}\,.
\end{equation}
The late-time value of $\langle\chi_{\pm}^{2}(t)\rangle$ in the zero temperature limit is then given by 
\begin{equation}
	\langle\chi_{\pm}^{2}(\infty)\rangle=\frac{c_{\pm}}{\pi m}\operatorname{Im}\Bigl[\mathfrak{I}_{1}^{(\pm)}\pm\mathfrak{I}_{2}^{(\pm)}\Bigr]\,.
\end{equation}

Similarly the late-time value of the expectation value $\langle p_{\pm}^{2}(t)\rangle$ in the zero temperature limit takes the form
\begin{align}
	\langle p_{\pm}^{2}(\infty)\rangle&=2mc_{\pm}\operatorname{Im}\int_{0}^{\infty}\!\frac{d\kappa}{2\pi}\;\biggl\{\frac{\kappa^{2}}{-\kappa^{2}-i\,2\gamma\kappa+\omega_{\pm}^{2}}-\frac{1}{\omega_{\pm}}\frac{\partial}{\partial\omega_{\pm}}\biggl[\frac{\pm\dfrac{\gamma}{\ell}\,\kappa^{2}\,e^{i\,\kappa\ell}}{-\kappa^{2}-i\,2\gamma\kappa+\omega_{\pm}^{2}}\biggr]\biggr\}\,.
\end{align}
We start with the integral
\begin{equation}
	\mathfrak{J}_{1}^{(\pm)}=\int_{0}^{\infty}\!d\kappa\;\frac{\kappa^{2}}{-\kappa^{2}-i\,2\gamma\kappa+\omega_{\pm}^{2}}\,,
\end{equation}
which apparently will diverge. Thus we replace the upper limit of the integral by a cutoff frequency $\Lambda$ to regularize the integral. We find, after regularization, $\mathfrak{J}_{1}^{(\pm)}$ is given by
\begin{align}\label{E:sdowuew}
	\mathfrak{J}_{1}^{(\pm)}=-\Lambda-\frac{i}{2\Omega_{\pm}}\Bigl[-\pi\bigl(\Omega_{\pm}-i\,\gamma\bigr)^{2}+2\bigl(\Omega_{\pm}^{2}-\gamma^{2}\bigr)\tan^{-1}\frac{\gamma}{\Omega_{\pm}}+2\Omega_{\pm}\gamma\,\ln\frac{\Omega_{\pm}^{2}+\gamma^{2}}{\Lambda^{2}}\Bigr]+\mathcal{O}(\frac{1}{\Lambda})\,.
\end{align}
This type of cutoff-dependent expression is often seen in a system interacting with a quantum-field environment, as the consequence of an infinite number of degrees of freedom associated with the field. The introduction of the cutoff scale is based on the understanding that any effective physical system or model has its range of validity. The cutoff frequency thus represents the highest energy scale permissible with the model. The leading term in \eqref{E:sdowuew}, which is linear in $\Lambda$, is of no relevance to us because it belongs to the real part of $\mathfrak{J}_{1}^{(\pm)}$. What is of our concern will be the term that is proportional to $\ln\Lambda$, which has a mild dependence on the cutoff scale, so it does not pose a serious issue. On the other hand $\mathfrak{J}_{2}^{(\pm)}$, defined by
\begin{align}
	\mathfrak{J}_{2}^{(\pm)}=-\frac{1}{\omega_{\pm}}\frac{\partial}{\partial\omega_{\pm}}\int_{0}^{\infty}\!d\kappa\;\frac{\pm\dfrac{\gamma}{\ell}\,\kappa^{2}\,e^{i\,\kappa\ell}}{-\kappa^{2}-i\,2\gamma\kappa+\omega_{\pm}^{2}}
\end{align}
is independent of the cutoff scale. The integral in $\mathfrak{J}_{2}$ is evaluated to be
\begin{align*}
	\int_{0}^{\infty}\!d\kappa\;\frac{\pm\dfrac{\gamma}{\ell}\,\kappa^{2}\,e^{i\,\kappa\ell}}{-\kappa^{2}-i\,2\gamma\kappa+\omega_{\pm}^{2}}&=-\frac{1-e^{i\,\Lambda\ell}}{\ell^{2}}+i\,\frac{\gamma}{2\Omega_{\pm}\ell}\bigl(\Omega_{\pm}-i\,\gamma\bigr)^{2}e^{+i\,(\Omega_{\pm}-i\,\gamma)\ell}\Bigl(\pi-i\,\operatorname{Ei}[-i(\Omega_{\pm}-i\gamma)\ell]\Bigr)\notag\\
	&\qquad\qquad\qquad+i\,\frac{\gamma}{2\Omega_{\pm}\ell}\bigl(\Omega_{\pm}+i\,\gamma\bigr)^{2}e^{-i\,(\Omega_{\pm}+i\,\gamma)\ell}\Bigl(\pi+i\,\operatorname{Ei}[+i(\Omega_{\pm}+i\gamma)\ell]\Bigr)\,,
\end{align*}
so $\mathfrak{J}_{2}^{(\pm)}$ is given by
\begin{align}
	\mathfrak{J}_{2}^{(\pm)}&=i\,\frac{\gamma^{2}}{\Omega_{\pm}^{2}\ell}-\frac{\gamma}{2\Omega_{\pm}^{3}\ell}\Bigl[+i\bigl(\Omega_{\pm}^{2}+\gamma^{2}\bigr)-\bigl(\Omega_{\pm}-i\,\gamma\bigr)^{2}\Omega_{\pm}\ell\Bigr]e^{+i\,(\Omega_{\pm}-i\,\gamma)\ell}\Bigl(\pi-i\,\operatorname{Ei}[-i(\Omega_{\pm}-i\gamma)\ell]\Bigr)\notag\\
	&\qquad\qquad+\frac{\gamma}{2\Omega_{\pm}^{3}\ell}\Bigl[-i\bigl(\Omega_{\pm}^{2}+\gamma^{2}\bigr)-\bigl(\Omega_{\pm}+i\,\gamma\bigr)^{2}\Omega_{\pm}\ell\Bigr]e^{-i\,(\Omega_{\pm}+i\,\gamma)\ell}\Bigl(\pi+i\,\operatorname{Ei}[+i(\Omega_{\pm}+i\gamma)\ell]\Bigr)\,.
\end{align}
Again we see $\mathfrak{J}_{2}^{(\pm)}$ is imaginary. Therefore we have $\langle p_{\pm}^{2}(\infty)\rangle$ in the zero temperature limit given by
\begin{equation}
	\langle p_{\pm}^{2}(\infty)\rangle=\frac{c_{\pm}m}{\pi}\operatorname{Im}\Bigl[\mathfrak{J}_{1}^{(\pm)}\pm\mathfrak{J}_{2}^{(\pm)}\Bigr]\,.
\end{equation}

\end{document}